\documentclass[twocolumn]{aastex63}

\usepackage{amsmath}

\received{December 31, 2019}
\revised{January 30, 2020}
\accepted{February 10, 2020}
\submitjournal{AJ}

\shorttitle{Intrinsic polarization in WR winds}
\shortauthors{Fullard et al.}
\graphicspath{{./}{figures/}}

\begin{document}

\title{A multi-wavelength search for intrinsic linear polarization in Wolf-Rayet winds}

\correspondingauthor{Andrew G. Fullard}
\email{andrew.fullard@du.edu}

\author[0000-0001-7343-1678]{Andrew G. Fullard}
\affiliation{Department of Physics and Astronomy, University of Denver, 2112 E. Wesley Ave, 80210, USA}
\author[0000-0003-3890-3400]{Nicole St-Louis}\affiliation{D\'epartement de physique, Universit\'e de Montr\'eal, C. P. 6128, succ. centre-ville, Montr\'eal (Qc) H3C 3J7}\affiliation{Centre de Recherche en Astrophysique du Qu\'ebec, Canada}
\author[0000-0002-4333-9755]{Anthony F. J. Moffat}
\affiliation{D\'epartement de physique, Universit\'e de Montr\'eal, C. P. 6128, succ. centre-ville, Montr\'eal (Qc) H3C 3J7}\affiliation{Centre de Recherche en Astrophysique du Qu\'ebec, Canada}
\author{Vilppu E. Piirola}
\affiliation{University of Turku, Finnish Centre for Astronomy with ESO, V\"ais\"al\"antie 20, 21500 Piikki\"o, Finland}
\author[0000-0001-7548-0190]{Nadine Manset}
\affiliation{CFHT Corporation, 65-1238 Mamalahoa Hwy, Kamuela, 96743, USA}
\author[0000-0003-1495-2275]{Jennifer L. Hoffman}
\affiliation{Department of Physics and Astronomy, University of Denver, 2112 E. Wesley Ave, 80210, USA}




\begin{abstract}
Wolf-Rayet stars have strong, hot winds, with mass-loss rates at least a factor of ten greater than their O-star progenitors, although their terminal wind speeds are similar. In this paper we use the technique of multiband linear polarimetry to extract information on the global asymmetry of the wind in a sample of 47 bright Galactic WR stars. Our observations also include time-dependent observations of 17 stars in the sample. The path to our goal includes removing the dominating component of wavelength-dependent interstellar polarization (ISP), which normally follows the well-known Serkowski law. We include a wavelength-dependent ISP position angle parameter in our ISP law and find that 15 stars show significant results for this parameter. We detect a significant component of wavelength-independent polarization due to electron scattering in the wind for 10 cases, with most WR stars showing none at the $\sim$0.05\% level precision of our data. The intrinsically polarized stars can be explained with binary interaction, large-scale wind structure, and clumping. We also found that 5 stars out of 19 observed with the Str\"omgren $b$ filter (probing the complex $\lambda$4600--4700 emission line region) have significant residuals from the ISP law and propose that this is due to wind asymmetries. We provide a useful catalogue of ISP for 47 bright Galactic WR stars and upper limits on the possible level of intrinsic polarization.

\end{abstract}

\keywords{stars: Wolf-–Rayet --- binaries: general --- methods: observational --- techniques: polarimetric --- surveys}


\section{Introduction\label{sec:intro}} 
Massive hot stars have high luminosities and as a result they drive strong winds via line-driving radiation pressure, mainly through UV photons interacting with ions in their hot wind. Those massive stars evolved to the cool part of the H-R diagram drive strong winds via radiation pressure mainly on dust grains (e.g. \citealt{lafon_mass_1991}). Above initial masses of $\sim$20 M$_\odot$, massive stars evolve into a classical Wolf-Rayet (cWR) stage of He-burning, with most of their outer H-rich envelopes removed by stronger winds in an intervening, relatively short LBV stage (or possibly an RSG stage for the least massive among them; \citealt{smith_mass_2014, smith_luminous_2017}). The extremely strong winds of cWR stars exceed those of their main-sequence (MS) O-star progenitors by at least an order of magnitude, even though their luminosities are rarely greater than those of their progenitors \citep{crowther_physical_2007}. The prime reason for this difference between MS and cWR stars is that cWR are 
near the Eddington limit. In most cases they also have high surface temperatures, from which the enhanced UV flux can drive strong winds due mainly to the large number of atomic transitions of iron in various ionization stages in the UV \citep{hillier_wc_1989}.

Another branch of stars with WR-like spectra are the most massive and luminous MS stars known, mostly of generic type WNLh or O/WNLh (with h sometimes replaced by ha or (h)). We include such stars if they are in the updated online general WR catalogue \citep{crowther_wolf-rayet_2015}. For convenience, we group cWR and these luminous H-rich stars under one designation, i.e.``WR''.

With typical mass-loss rates of $10^{-5}$ M$_{\odot}$/yr and terminal velocities of 2000 km/s, WR winds are optically thick out to about 2 R$_*$ (where R$_*$ is the hydrostatic core radius) and optically thin beyond this \citep{hamann_galactic_2019}. The outer, thin part is stratified, with emission lines of higher ionization formed closer to the hotter lower boundary and lines of lower ionization formed further out, although with a degree of overlap between the ionization groups (e.g. \citealt{hillier_wc_1989}). The inner thick wind remains essentially unobservable, making it impossible to directly probe the key stellar properties at R$_*$. But one can nevertheless get a reasonable indirect handle on these parameters by modelling the emerging emission-line spectrum \citep{hamann_galactic_2019}. Another technique  is to track the trajectories of inhomogeneities in the outer, observable wind, such as clumps and co-rotating interaction regions (CIRs) as seen in O stars, which have their origin in the inner wind region, if not at R$_*$ itself  \citep{ramiaramanantsoa_brite_2018, ramiaramanantsoa_chaotic_2019}.

Another factor affecting WR winds is 
the rotation of the underlying star; rapid rotation is likely an important element in creating long-duration gamma-ray bursts (LGRBs \citealt{woosley_supernova_2006}). Rapid rotation of some WR stars has been inferred by  \citet{harries_spectropolarimetric_1998} and others using line depolarization. In this model, the flattened wind leads to higher polarization in continuum light, which mostly arises from near the base of the optically thin wind. This is accompanied by less polarization of lines with lower ionization states as they are formed further out in the wind where there are fewer free electrons off which to scatter. The scattering of light by free electrons (or ions, to a much lesser degree) leads to polarization in an {\em asymmetric} wind, whereas a spherically symmetric wind will show no net polarization for any lines or continuum. However, \citet{stevance_probing_2018} found that they could not rule out the presence of rapid rotation using the lack of a line effect as the sole diagnostic.

Any Galactic polarimetric measurement contains an interstellar polarization (ISP) contribution due to scattering of starlight by aligned dust grains in the interstellar medium. There are multiple ways to extract the intrinsic stellar polarization from such measurements \citep[e.g.,][]{quirrenbach97constraints}.
For example, observing the target using spectropolarimetry allows one to use non-variable line polarization to estimate and then subtract the ISP contribution \citep{harrington_intrinsic_1968}. Since the ISP does not change rapidly with time, fitting observed polarization variability using models such as that of \citet{brown_polarisation_1978} can also recover the intrinsic polarization component. We use a third method, characterizing the ISP by obtaining multi-wavelength broadband polarimetric observations and simultaneously fitting the empirical Serkowski law describing the ISP behavior
\citep{serkowski_wavelength_1975, wilking_wavelength_1980, whittet_systematic_1992} along with a wavelength-independent,  constant level of polarization assumed intrinsic to the star.
This achieves both characterization of the ISP and identification of any significant 
continuum polarization caused by free-electron scattering in a flattened wind. In this work, we apply this Serkowski + constant fit method to  continuum-dominated  polarimetry of a sample of 47 Galactic WR stars. Our observations (taken between 1989 and 1991) used broadband $UBVRI$ filters, as well as a Str\"omgren $b$ filter in some cases to isolate the WR emission-line complex at $\sim$4650\AA. We present our data in Section 2. In Section 3, we discuss in more detail the cases of six stars that show polarimetric time variability. Our fits to the ISP and intrinsic polarization are the subject of Section 4. Finally, we discuss our results in Section 5 and conclude in Section 6.  


\section{Data}
We obtained our multiband polarimetry 
in two observing runs, one in the North at 
the 1.25m Crimean Observatory telescope in Sept 1989, the other in the South at the 1.5m 
ESO/La Silla telescope in May 1991. Both these telescopes were equipped with a simultaneous 5-channel polarimeter designed and built by V. Piirola \citep{piirola_double_1973, piirola_simultaneous_1988, korhonen_polarization_1984}. At La Silla the photo-tube in the $I$-band was malfunctioning, so we replaced this filter with a medium band Str\"omgren $b$ filter to simulate partial spectropolarimetry. Only the brightest stars had enough flux to give useful data in this narrower filter. Fortunately, the lack of $I$-band data proved not to be a major handicap when we fit the data as a function of wavelength (Section~\ref{sec:fitting}). We calibrated the polarization angles in each filter using standard polarized stars.  We also observed unpolarized standard stars to eliminate the 
instrumental polarization (which was very small, less than 0.01\%\ in all bands).

Table~\ref{tab:starlist} lists the stars we observed, along with their $V$ magnitudes, spectral types, binary status, and, if applicable, their periodicities (including 
those due to non-binary variation), all taken from the online WR catalogue of \citet{crowther_wolf-rayet_2015} unless stated otherwise.



\startlongtable
\begin{deluxetable*}{lllllcDc}
\tablecaption{Basic parameters for stars included in our sample. $M_V$ taken from the SIMBAD database. Spectral type and binary status are taken from \citet{crowther_wolf-rayet_2015} unless otherwise noted. Periods are taken from \citet{van_der_hucht_viith_2001, van_der_hucht_new_2006} unless otherwise noted. \label{tab:starlist}}
\tablehead{
\colhead{WR} & \colhead{HD} & \colhead{Alt ID} & \colhead{$M_V$}  & \colhead{Spectral type} & \colhead{Binary status} & \multicolumn2c{Period (d)} & \colhead{Ref.}
}
\startdata
\cutinhead{ESO/La Silla}
6\tablenotemark{ac}   & 50896 & EZ CMa & 6.91 & WN4b                                    & CIR?  & 3.77  & &\nodata   \\
8   & 62910 &  & 10.10 & WN7o/CE                                 & SB1 &  38.4 & &\nodata  \\
9   & 63099 & V443 Pup & 10.50 & WC4 + O7                                & SB2  & 14.305   & & 1 \\
14  & 76536 &  & 8.80 & WC7+?                                     & SB1  &  2.42 & &\nodata  \\
16\tablenotemark{a}  & 86161 & V396 Car & 8.36 & WN8h                                    & \nodata  &  \nodata & & \nodata \\
21  & 90657 & V398 Car & 9.65 & WN5o + O4-6                             & SB2  & 8.25443  & & 2 \\
22\tablenotemark{a}  & 92740 & V429 Car & 6.42  & WN7h + O9 V-III                         & SB2  & 80.336  & &\nodata  \\
23  & 92809 &  & 9.03 & WC6                                     & \nodata  & \nodata& &\nodata  \\
24  & 93131 &  & 6.48 & WN6ha                                   & \nodata  & \nodata  & &\nodata \\
25  & 93162 &  & 8.80 & O2.5 If* / WN6 + O                      & SB2  & 207.85 & & 3   \\
40\tablenotemark{a} & 96548 & V385 Car & 7.70 & WN8h                    & \nodata  &  \nodata  & &\nodata \\
42\tablenotemark{bc}  & 97152  & V431 Car & 8.07 & WC7 + O7V                               & SB2  & 7.8912 & & 4  \\
43\tablenotemark{a}  & 97950 & NGC3603abc & 9.03 & a=A1: WN6ha + WN6ha        & SB2   & 3.7724  & & 5   \\
  & & &  & c=C: WN6ha + ?         & SB1   &  8.89   & & 5  \\
46  & 104994 & DI Cru & 10.93 & WN3bp                                   & ?  & 0.28-0.33& & 6  \\
48\tablenotemark{a} & 113904 & * $\theta$ Mus  & 5.53 & WC6 + O6-7V( + 09.7Iab)         & SB1  & 19.1375 & & 7  \\
52  & 115473 & & 9.00 & WC4                                     & \nodata &\nodata & &\nodata  \\
57  & 119078 & & 9.40 & WC8                                     &\nodata & \nodata & &\nodata \\
69  & 136488 & & 9.10 & WC9d + OB                              & SB2 & 2.293 & &\nodata  \\
71\tablenotemark{a}  & 143414 & LT TrA & 10.10 & WN6o                                    & SB2? & 7.69 & &\nodata \\
78  & 151932 & V919 Sco & 6.51 & WN7h                                    &\nodata & \nodata & &\nodata \\
79\tablenotemark{bc} & 152270 & & 6.59 & WC7 + O5-8                              & SB2  & 8.8911  & & 4 \\
86  & 156327 & V1035 Sco & 9.32 & WC7 (+ B0III-I)                         & VB   &   0.1385   & &\nodata    \\
90  & 156385 & & 6.92 & WC7                                     &\nodata  & \nodata & &\nodata  \\
92  & 157451 & & 10.20 & WC9                                     &\nodata  & \nodata & &\nodata  \\
103\tablenotemark{a} & 164270 & V4072 Sgr & 8.74 & WC9d + ?                                   & SB1  & 1.7556 & &\nodata  \\
108 & 313846 & & 9.89 & WN9ha                                   &\nodata  & \nodata & &\nodata  \\
110 & 165688 & & 9.87 & WN5-6b                                  & CIR?  & 4.08 & & 8  \\
111\tablenotemark{a} & 165763 & & 7.82 & WC5                                     &\nodata  & \nodata & &\nodata  \\
113\tablenotemark{a} & 168206 & CV Ser & 9.10 & WC8d + O8-9IV                           & SB2  & 29.700  & & 9  \\
123 & 177230 & V1402 Aql & 11.12 & WN8o                                    & SB1? &  2.3940 & &\nodata  \\
\cutinhead{Crimean Observatory}
1   & 4004 & V863 Cas  & 10.14 & WN4b                                    & SB1? & \nodata & &\nodata \\
3   & 9974 & & 10.69   & WN3ha                                   & SB2  & 46.85 & &\nodata \\
127 & 186943 & QY Vul & 10.69 & WN5o + O8.5V                            & SB2  & 9.5550  & & 10 \\
128 & 187282 & QT Sge & 10.51 & WN4(h)                                 & SB2? &   3.56  & &\nodata \\
133\tablenotemark{b} & 190918 & V1676 Cyg & 6.75 & WN5o + O9I                              & SB2  & 112.4    & &\nodata \\
134\tablenotemark{b} & 191765 & V1769 Cyg & 8.08 & WN6b                                    & CIR & 2.255  & & 11   \\
135 & 192103 & V1042 Cyg & 8.11 & WC8                                     & \nodata & \nodata& &\nodata  \\
136 & 192163 & V1770 Cyg & 7.50 & WN6b(h)                                 & SB1? & 4.554 & &\nodata \\
137 & 192641 & V1679 Cyg & 7.91 & WC7pd + O9                             & SB2  & 4766  & & 12   \\
138 & 193077 & & 8.01 & WN5o + B?                               & SB2  & 1538   & & 13   \\
139\tablenotemark{bc} & 193576 & V444 Cyg & 8.00 & WN5o + O6V-III                          & SB2  & 4.212454 & & 14 \\
140 & 193793 & V1687 Cyg & 6.85 & WC7ed + O5.5fc                          & SB2  & 2900 & & 15 \\
141\tablenotemark{bc} & 193928 & V2183 Cyg & 9.78 & WN5o + O5V-III                         & SB2  & 21.6895 & &\nodata  \\
148 & 197406 & V1696 Cyg & 10.30 & WN7ha + O4-6V                           & SB2  & 4.317336& & 16  \\
153 & 211853 & GP Cep & 9.00 & a1: WN6o/CE + O3-6 & SB2  & 6.6887  & & 17 \\
 &  & &  & a2: B0:I + B1:V-III & SB2  & 3.4663 & & 17  \\
155\tablenotemark{bc} & 214419 & CQ Cep & 8.80 & WN6o + O9II-Ib    & SB2  & 1.6412436 & & \nodata  \\
157 & 219460B & & 10.75 & WN5o (+ B1II)   & VB   & 1.7860 &  &\nodata
\enddata
\tablenotetext{a}{Denotes systems with 2--5 observations.}\tablenotetext{b}{Denotes systems with more than 5 observations.}\tablenotetext{c}{Denotes systems for which our data have been previously published.}
\tablerefs{1: Spectral Type~\citet{bartzakos_magellanic_2001}, 2: Spectral Type~\citet{fahed_colliding_2012}, 3: Period~\citet{gamen_first_2006}, 4: Period~\citet{hill_modelling_2000}, 5: Period~\citet{schnurr_very_2008}, 6: Period~\citet{marchenko_puzzle_2000}, 7: Period~\citet{hill_modelling_2002}, 8: Binary status~\citet{st-louis_systematic_2009}, 9: Period~\citet{hill_modelling_2018}, 10: Period~\citet{de_la_chevrotiere_spectroscopic_2011}, 11: Period~\citet{aldoretta_extensive_2016}, 12: Period~\citet{lefevre_spectroscopic_2005}, 13: Period~\citet{annuk_long_1990}, 14: Period~\citet{eris_2007_2011}, 15: Period~\citet{williams_variable_2019}, 16: Period~\citet{munoz_wr_2017}, 17: Period~\citet{demers_quadruple_2002}}
\end{deluxetable*}

\section{Mean polarization of time-dependent observations\label{sec:binaryfit}}
For systems with multiple observations, we require a single mean polarization value per band so that we can calculate the constant intrinsic and ISP components. We obtained these mean values in one of two ways, depending on the system and number of observations. In the case of binaries with known orbital periods, we fitted theoretical binary polarization models to our data in each waveband and took the resulting constant $q$ and $u$ values to represent a ``systemic mean" polarization for the system. To fit the models, we used previously derived binary parameters from Table~\ref{tab:binaryparameters}. These fits also allowed us to derive new physical parameters for WR 133; see $a)$ below. For single stars, we took an uncertainty-weighted mean of the polarization measurements in each band. 
Table~\ref{tab:snapshotdata} tabulates these mean $UBVRIb$ polarimetric values and uncertainties; we discuss individual cases in the subsections below.

\begin{deluxetable*}{lccccccc}
\tablecaption{Extant estimated parameters for systems with time-dependent data that were fit in Section~\ref{sec:binaryfit}. 
\label{tab:binaryparameters}}
\tablehead{WR  & \colhead{$E_0~(\mathrm{HJD})$}  & \colhead{$P (\mathrm{d})$}  & \colhead{$e$}  & \colhead{$i~(\degr)$}                         & \colhead{$\Omega~(\degr)$}  & \colhead{$\omega_{\mathrm{WR}}~(\degr)$} & Ref.}
\startdata
133 & $2447420.5\pm0.036$ & $112.4\pm0.02$   & $0.39\pm0.007$ & \nodata  & \nodata    & $18.9\pm0.0107$  & 1\\
134 & \nodata  & $2.255\pm0.0008$   & \nodata     & \nodata & \nodata   & \nodata     & 2  \\
139 & $2441164.311\pm0.007$   & $4.212454\pm0.000004$   & 0.00    & $80.8\pm1.6$ & $-41.8\pm3.8$ & \nodata   & 3    \\
141 & $2448840.80\pm0.002$ & $21.6895\pm0.00003$ & 0.00    & $68\pm12$    & $103\pm25$    & \nodata  & 4   
\enddata
\tablerefs{1: \citet{underhill_study_1994, robert_polarization_1989}, 2: \citet{aldoretta_extensive_2016}, 3: \citet{eris_2007_2011}{($E_0$, $P$)}; \citet{st._-louis_polarization_1993}{($i$, $\Omega$)}, 4: \citet{marchenko_wolf-rayet_1998}}
\end{deluxetable*}

\begin{deluxetable*}{lcccccDDDD}
\tablecaption{Mean polarization data for all our targets, calculated as described in Section~\ref{sec:binaryfit}.
\label{tab:snapshotdata}}
\tablehead{
\colhead{WR} & \colhead{Variability} & \colhead{Mean} & \colhead{Obs. count} & \colhead{HJD} & \colhead{Band} & \multicolumn2c{$q$ (\%)} & \multicolumn2c{$\sigma_q$ (\%)} & \multicolumn2c{$u$ (\%)} & \multicolumn2c{$\sigma_u$ (\%)}\\
\colhead{} & \colhead{} & \colhead{} & \colhead{} & \colhead{2,440,000+} & \colhead{} & \multicolumn2c{} & \multicolumn2c{} & \multicolumn2c{} & \multicolumn2c{}
}
\startdata
\decimals
1   & SB1? & S  & 1  & 7768.5470 & $U$ & -5.713 & 0.101 & -1.148 & 0.204  \\
    &      &    &    &           & $B$ & -6.243 & 0.053 & -1.383 & 0.086  \\
    &      &    &    &           & $V$ & -6.442 & 0.092 & -1.423 & 0.104 \\
    &      &    &    &           & $R$ & -5.748 & 0.047 & -1.360 & 0.059  \\
    &      &    &    &           & $I$ & -5.122 & 0.065 & -1.163 & 0.059  \\
\enddata
\tablecomments{Table~\ref{tab:snapshotdata} is published in its entirety in the machine-readable format. A portion is shown here for guidance regarding its form and content.}
\end{deluxetable*}

\textbf{a) WR 133} This is a binary WN5o + O9I system. Its observed polarimetric data are presented in the appendix, Table~\ref{tab:wr133data}. To calculate its systemic mean polarization, we followed \citet{moffat_luminous_1998}, fitting both $q$ and $u$ simultaneously with an analytical polarization model for elliptical binary orbits derived from \citet{brown_effect_1982}, corrected by \citet{simmons_interpretation_1984} and modified for an extended source of scatterers (see \citealt{robert_photometry_1992}). The model equations are 
\begin{equation}\label{eqn:qellipse}
    q = q_0 + \Delta q \cos\Omega-\Delta u\sin\Omega,
\end{equation}
\begin{equation}\label{eqn:uellipse}
    u = u_0 + \Delta q \sin\Omega+\Delta u\cos\Omega,
\end{equation}
where
\begin{equation}\label{eqn:deltaq}
    \Delta q = -\tau_3[(1 + \cos^2 i) \cos 2\lambda - \sin^2 i]
\end{equation}
and
\begin{equation}\label{eqn:deltau}
    \Delta u = -2\tau_3\cos i - \sin2\lambda\,.
\end{equation}
The parameters $q_0$ and $u_0$, which we adopt as our systemic mean values, represent the interstellar (plus any constant intrinsic) polarization. As usual, $\Omega$ is the rotation of the line of nodes on the sky counter-clockwise from the north and $i$ is the orbital inclination with respect to the line of sight. The quantity $\lambda$ is defined by $\lambda = \nu + \omega_{\mathrm{WR}} + \pi/2$, where $\nu$ is the true anomaly and $\omega_{\mathrm{WR}}$ is the argument of periastron for the WR star. Finally, $\tau_3$ is given by $\tau_3 = \tau_*(a/r)^\gamma$, with $\tau_*$ representing the mean optical depth, $a$  the mean orbital separation, and $r$  the instantaneous separation. The parameters $a$ and $r$ are related by
\begin{equation}\label{eqn:aoverr}
    a/r=[1+e\cos(\lambda-\lambda_p)]/(1-e^2)\,,
\end{equation}
where $e$ is the orbital eccentricity and $\lambda_p$ is the periastron passage, with $\lambda_p = \omega_{\mathrm{WR}} + \pi/2$. In the expression for $\tau_3$, $\gamma$ is a power index that reflects the actual free-electron density around the WR star between two plausible extremes: $\gamma$ = 1 for a uniformly ionized wind and $\gamma$ = 2 for an idealized global point source of scatterers. This means that the free electrons in the WR wind are located tightly around the WR star so that we can ignore any extension in radius. 

WR 133 was also observed polarimetrically by~\citet{robert_polarization_1989}. We used their blue single-filter broadband data in the 0.6--0.9 phase region to improve the overall fit, treating this source as though it was simply another observed band with its own $q$ and $u$ zero points to be fitted. We discarded the~\citet{robert_polarization_1989} zero-point values because their data were not observed with the same instruments as ours.

\begin{figure*}
\gridline{\fig{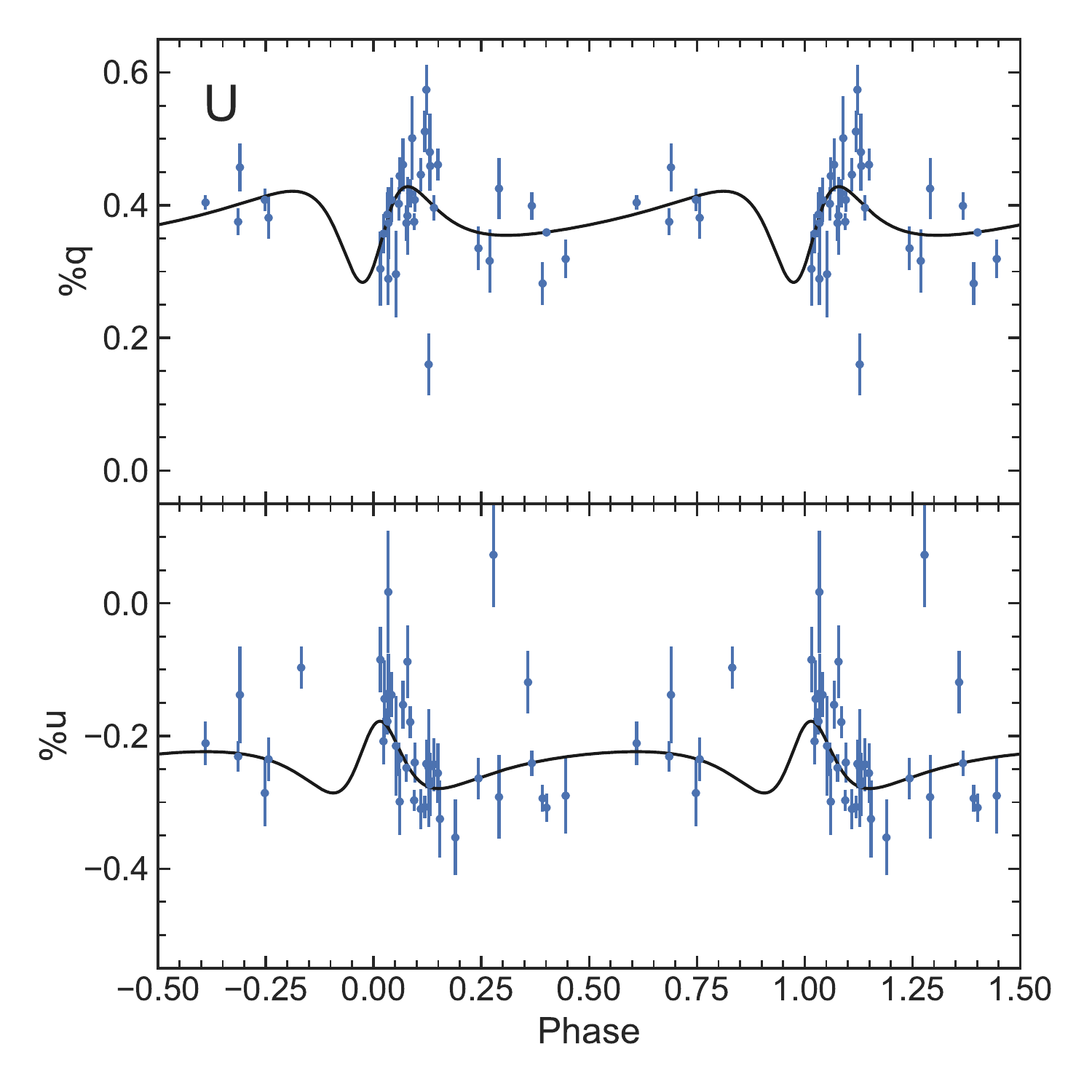}{0.33\textwidth}{(a)}
\fig{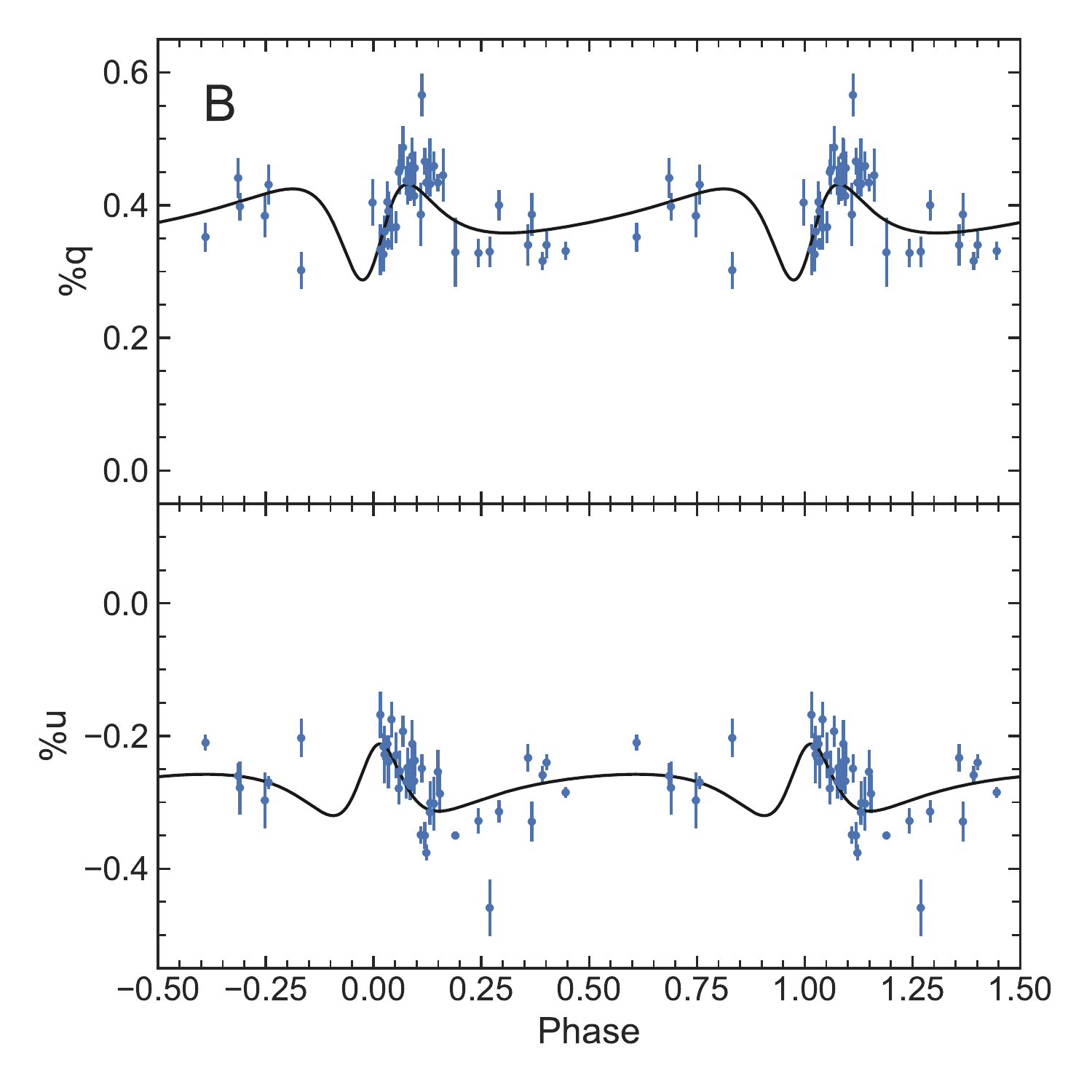}{0.33\textwidth}{(b)}
\fig{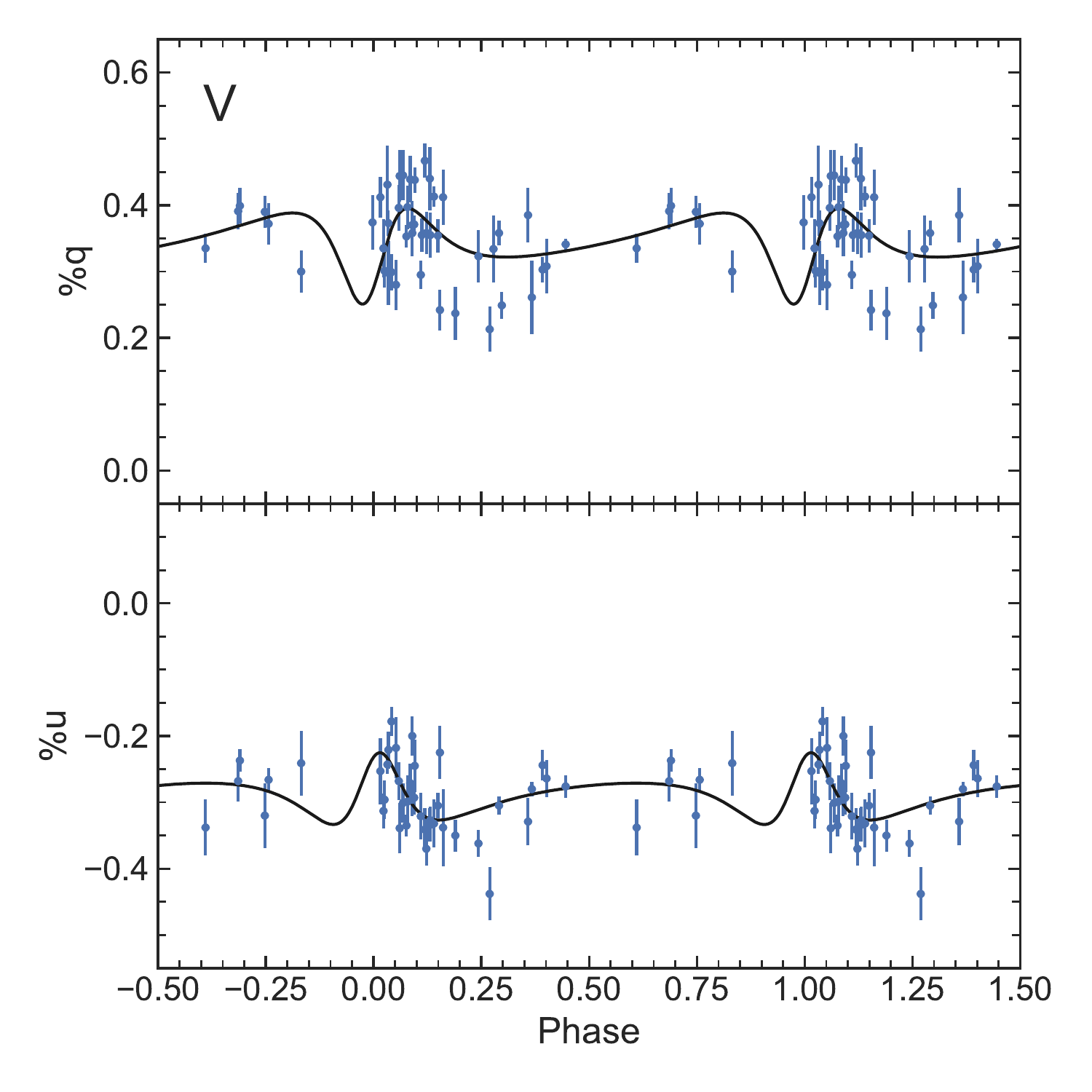}{0.33\textwidth}{(c)}}
\gridline{\fig{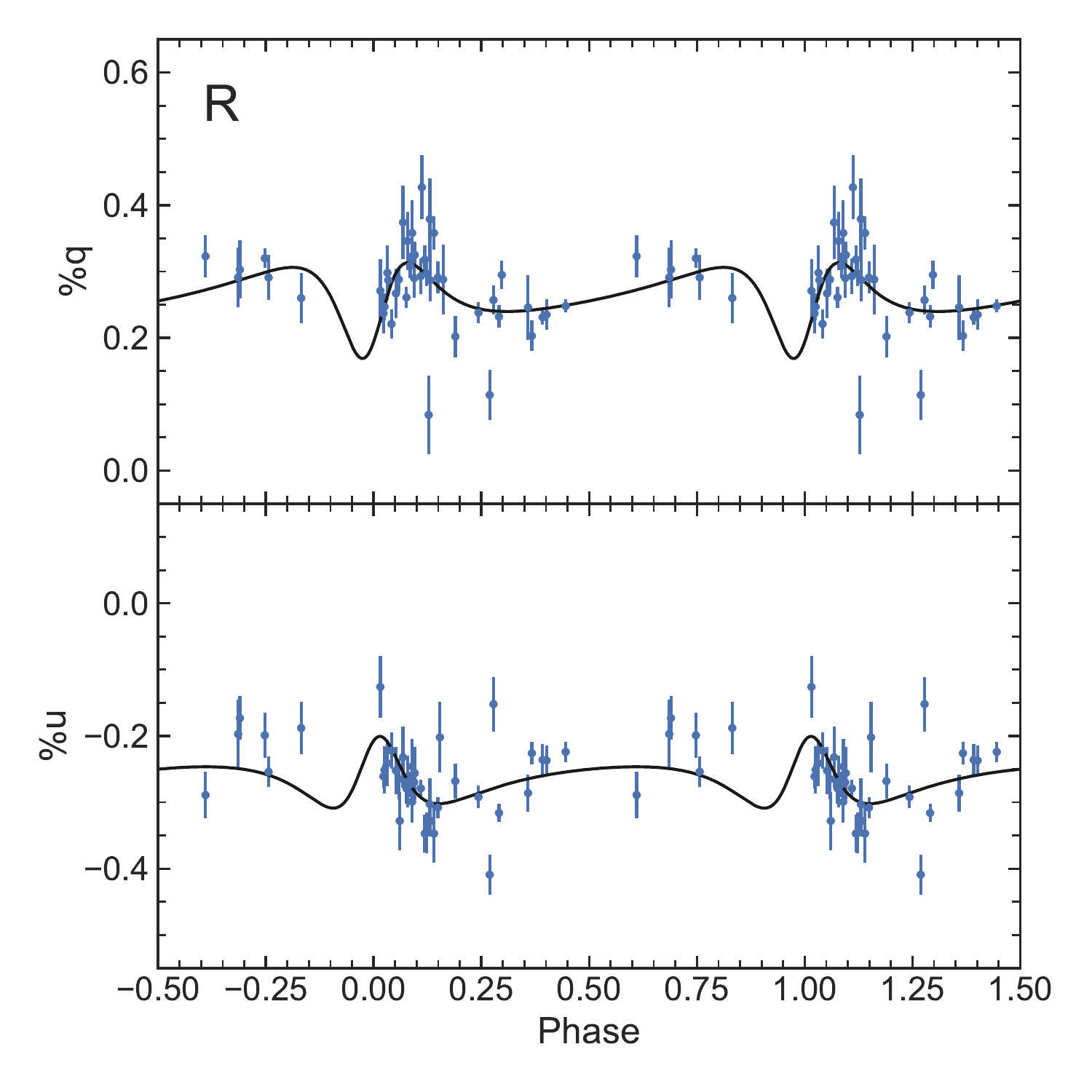}{0.33\textwidth}{(d)}
\fig{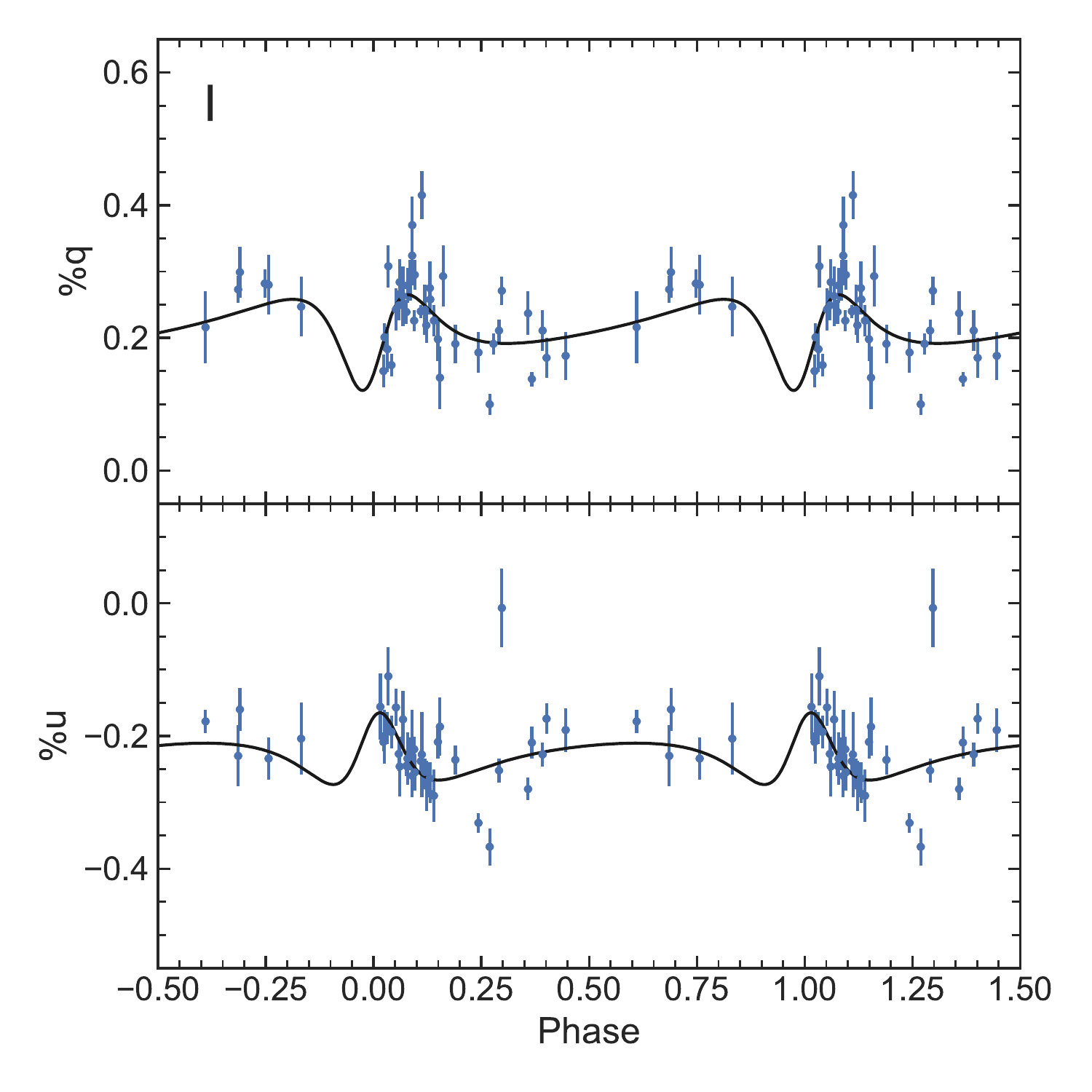}{0.33\textwidth}{(e)}
\fig{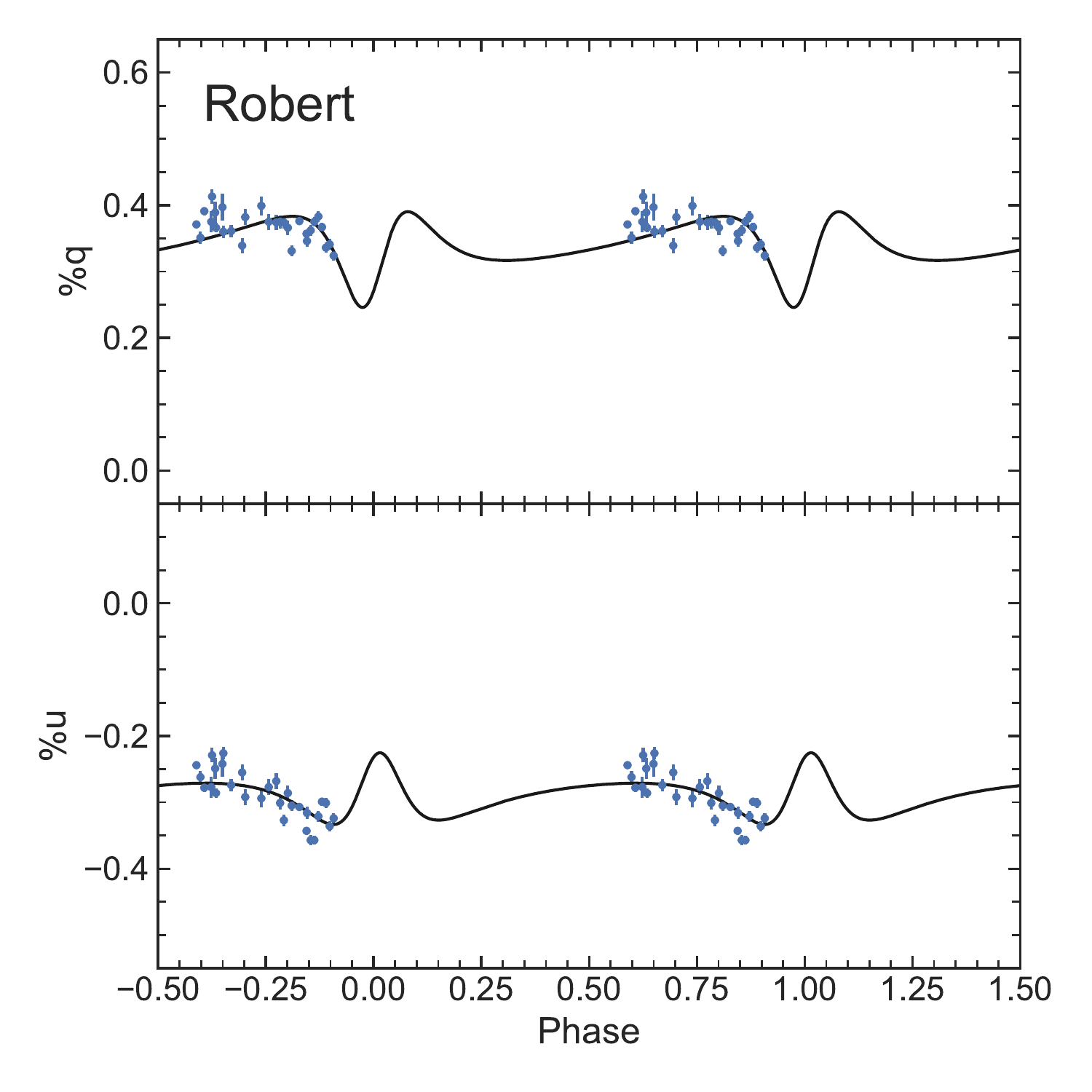}{0.33\textwidth}{(f)}}
\caption{$UBVRI$ filter data and orbital fits for WR 133 (Section~\ref{sec:binaryfit}\textbf{a}). Panels (a) through (e) correspond to $U$ through $I$ filters. Panel (f) displays data from \citet{robert_polarization_1989} for comparison. The black lines represent our fit to the data using equations~\ref{eqn:qellipse} and~\ref{eqn:uellipse}. The data presented in this figure are available in the Appendix, Table~\ref{tab:wr133data}.}
\label{fig:wr133data}
\end{figure*}

The polarization in WR stars is caused by electron scattering in the hot, ionised outflow, and as a consequence we expect it to be largely wavelength-independent. Therefore, for WR 133 we kept all parameters the same for each band, except for the $q$ and $u$ zero points ($q_0$ and $u_0$), then fitted all bands simultaneously in $q$ and $u$. We phased the data using the published ephemeris for the system (listed in Table~\ref{tab:binaryparameters}). We fixed $e$, $P$ and $\Omega$ using the estimates from \citeauthor{underhill_study_1994} (\citeyear{underhill_study_1994}; Table~\ref{tab:binaryparameters}). Lastly, we carried out the fit minimizing the uncertainty-weighted $\chi^2$ values as a function of $q_0$, $u_0$, $\Omega$, $\tau_3$, and $i$ with \textsc{lmfit} \citep{newville_lmfit:_2014}. We used the least-squares Trust Region Reflective method with Huber loss function to provide a robust method of dealing with outliers. We found that fixing $\gamma=1$ provided the best fit as measured by a Kolmogorov-Smirnov test of the Studentized residuals compared to a Gaussian distribution with $\mu=0$ and $\sigma^2=1$, though the data are not complete enough to reliably discriminate between $\gamma=1$ or 2. The systemic mean polarization values for each band are presented in Table~\ref{tab:snapshotdata}, and we list the fitted orbital parameters in Table~\ref{tab:wr133params}. The fits are displayed in Figure~\ref{fig:wr133data}.

Given our fitted value for the inclination, $i=115.9\degr\pm7.3\degr$, we attempted to calculate the masses of the components using the $M\sin^3 i$ values provided by \citet{underhill_study_1994}. We derived $M_O = 1.12\, $M$_\sun$ and $M_{WR} = 0.55\, $M$_\sun$, unrealistically low masses for both spectral types. Using the polarization-derived orbital parameter confidence intervals from \citet{wolinski_confidence_1994}, we find that our $\sigma_P/A$ metric is approximately 0.6, where $\sigma_P \approx 0.038\%$ is the average uncertainty of our polarization measurements and $A = (|q_{max} - q_{min}| + |u_{max} - u_{min}|) / 4 = 0.063\%$ describes the amplitude of the polarization variation fit. Using Fig. 5 from \citet{wolinski_confidence_1994}, we estimate the critical value of $i$ as $\sim70\degr$ or $\sim110\degr$, for which the upper limit of the possible inclination reaches $0\degr$ or $180\degr$, respectively. Thus 
our fitted inclination is more properly expressed as $i = 115.9\degr^{+64.1\degr}_{-7.3\degr}$. This unfortunately makes it difficult to derive further parameters of interest from our inclination angle with any confidence. Given the expected inclination range of 15--30$\degr$ \citep{underhill_study_1994}, Fig. 5 of \citet{wolinski_confidence_1994} suggests that given our current estimate for $A$, measurement uncertainties of less than $\sim0.0008\%$ are required to verify this small inclination angle polarimetrically.

Under the assumption that our $\tau_*$ value and the orbital separation values from \citet{underhill_study_1994} are correct, we provide an estimate of the mass-loss rate $\dot{M}$  using the following equation from \citet{moffat_luminous_1998} (see also \citealt{st-louis_polarization_1988}):

\begin{multline}
    \dot{M}_{WR} /2\times10^{-5}M_\odot ~\mathrm{yr}^{-1} = \\ \frac{\tau_*(v_\infty / 2000~\mathrm{km~ s}^{-1})(a/0.5~\mathrm{AU})}{0.0016 (f_c/0.6)(\alpha /0.5)}
\end{multline}

\noindent where $f_c$ is the fraction of the total light from the companion star, $\alpha$ is the number of scattering electrons per nucleon, $a$ is the mean orbital separation and $v_\infty$ is the WR terminal wind velocity. We adopt $v_{\infty}=1535$ km s$^{-1}$ from \citet{niedzielski_kinematical_2002}, $\alpha=0.5$ for fully ionized He, and calculate $f_c = I_O/(I_{WR} + I_O) = 10^{-6.55/-2.5} / (10^{-4/-2.5} + 10^{-6.55/-2.5}) = 0.913$ using absolute magnitudes from \citet{bowen_far_2008} and \citet{crowther_physical_2007} for the O and WR stars respectively. We adopt $a=1.154$ AU from \citet{underhill_study_1994}. This results in a low mass-loss rate of $\dot{M}_{WR} = 6.52\pm0.6\times10^{-6}M_\odot~\mathrm{yr}^{-1}$. This is within the upper limit reported by \citet{st-louis_polarization_1988}, and provides a tighter constraint for this system.


\begin{deluxetable}{lc}
\tablecaption{Fitted binary parameters for WR 133 (Section~\ref{sec:binaryfit}\textbf{a}).\label{tab:wr133params}}
\tablehead{\colhead{Parameter} & Value }
\startdata
$i$ (\degr)      & 115.9 $\pm$ 7.3\phm{0}      \\
$\Omega$ (\degr) & 162.4 $\pm$ 5.4\phm{0}      \\
$\tau_*$  & 4.48 $\pm$ 0.93 $\times 10^{-4}$  \\
$\dot{M}_{WR}~(M_\odot~\mathrm{yr}^{-1})$   & $6.52\pm0.6\times10^{-6}$
\enddata
\tablecomments{Based on the uncertainty analysis by \citet{wolinski_confidence_1994}, the $i$ presented here is a lower limit ($115.9\degr < i < 180\degr$).}
\end{deluxetable}

\textbf{b) WR 139 and WR 141} Although these data were previously published by \citeauthor{marchenko_wolf-rayet_1994} (\citeyear{marchenko_wolf-rayet_1998}; WR 139) and \citeauthor{st._-louis_polarization_1993} (\citeyear{st._-louis_polarization_1993}; WR 141), these authors did not provide the fit parameters $q_0$ and $u_0$. We therefore recalculated the fits to recover the systemic mean values. Since these binaries both have circular orbits, the elliptical prescription is not appropriate, so we fitted their data with circularized versions of equations~\ref{eqn:qellipse} and~\ref{eqn:uellipse}, where $\lambda = 2\pi\phi$ in equations~\ref{eqn:deltaq} and~\ref{eqn:deltau}, and $\phi$ is the orbital phase. Also, because $a=r$ for a circular orbit, $\tau_3 = \tau_*$.

For WR 139, we did not fit the data in the region between phases 0.4--0.6 because of its strong departure from the simple model due to eclipse effects \citep{st._-louis_polarization_1993}.
The resulting binary parameters we found for both systems are the same within uncertainties as those previously published, so we do not present them here.



\textbf{c) WR 134} This object has not been shown to have a luminous binary companion. Instead, the wind of WR 134 probably features rotating CIRs that come and go with a coherence timescale of about 40 days \citep{aldoretta_extensive_2016}. Therefore the binary models we used in \textbf{a)} and \textbf{b)} are not appropriate to describe its polarization variability. Instead, we phased our data to the period given in \citeauthor{aldoretta_extensive_2016} (\citeyear{aldoretta_extensive_2016}; Table~\ref{tab:binaryparameters}) and took an uncertainty-weighted mean in each band to represent the mean polarization.
We present the filter data in Figure~\ref{fig:wr134data}. Its observed polarimetric data are presented numerically in the Appendix, Table~\ref{tab:wr134data}. The $UBVRI$ mean values are presented in Figure~\ref{fig:wr134data} f) to better display the periodic behavior of the system. This periodic behavior has been seen in polarimetric data by \citet{morel_2.3_1999}. However, in contrast to the \citeauthor{morel_2.3_1999} results, our $u$ data lack a clear periodicity. This may be related to the coherency timescale of the wind structures, or a different location of the structures in the wind. Our $q$ data appear to phase well with the \citet{aldoretta_extensive_2016} period, suggesting that the period is related to a permanent feature of the star, such as its rotation rate. 

\begin{figure*}
\gridline{\fig{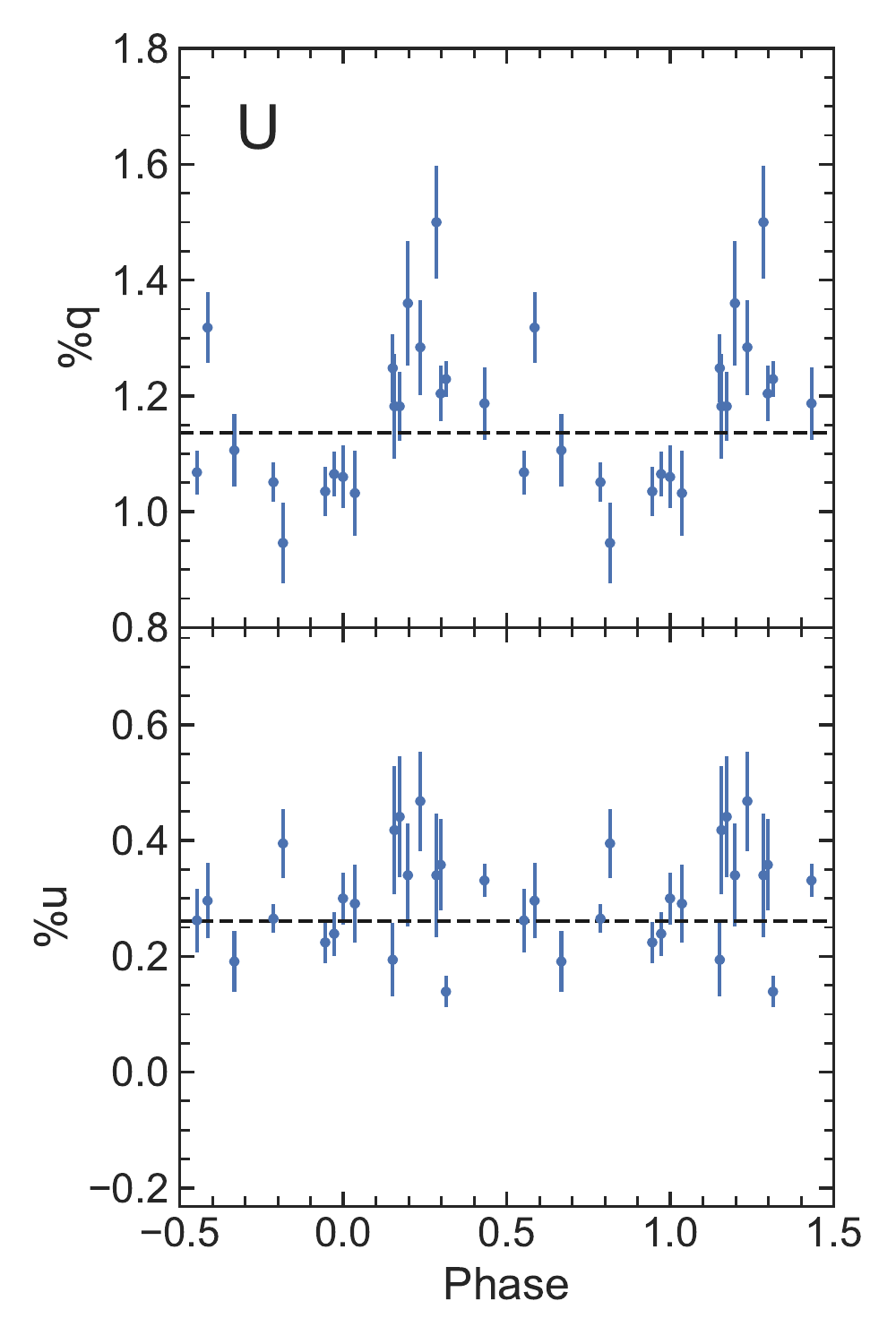}{0.33\textwidth}{(a)}
\fig{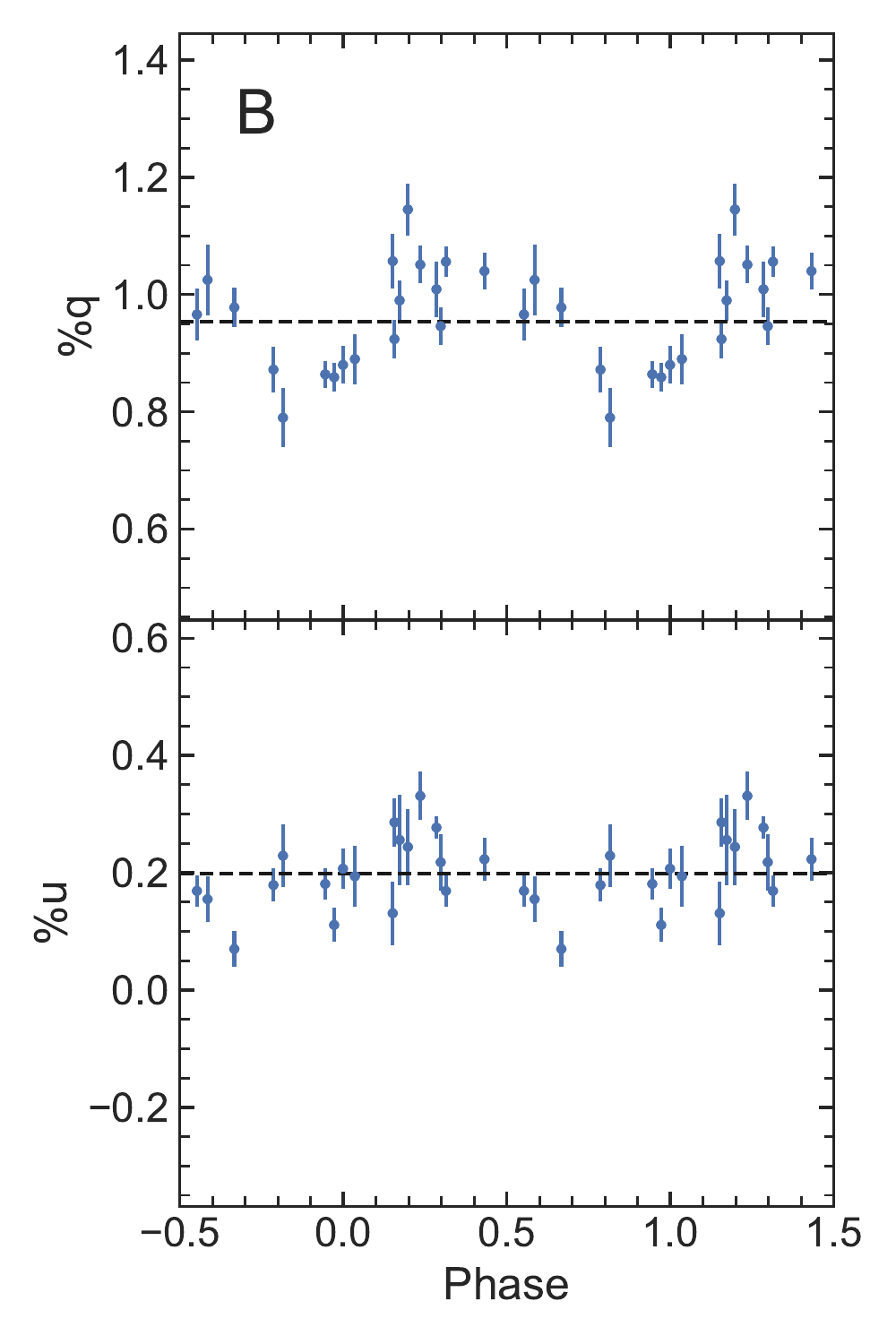}{0.33\textwidth}{(b)}
\fig{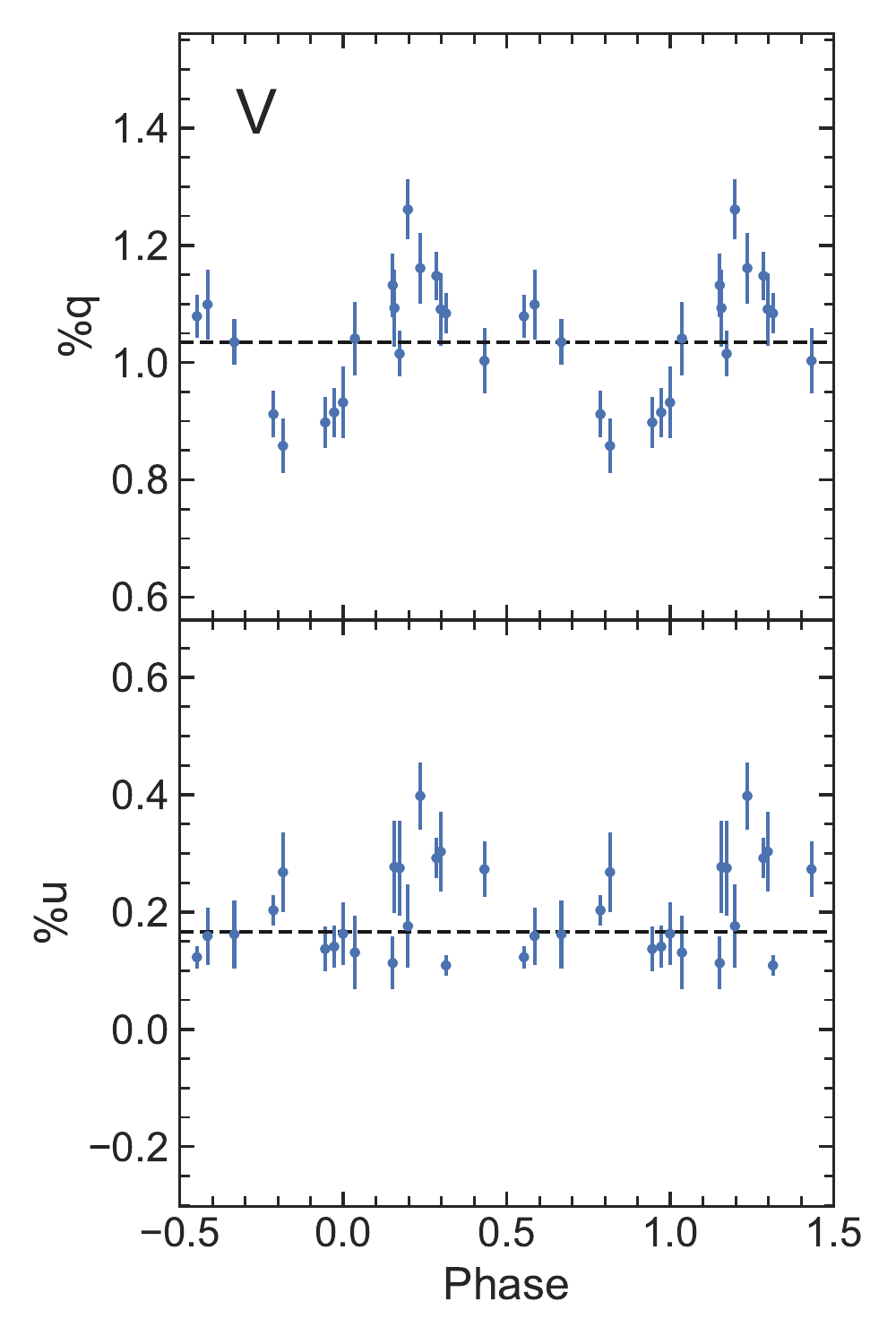}{0.33\textwidth}{(c)}}
\gridline{\fig{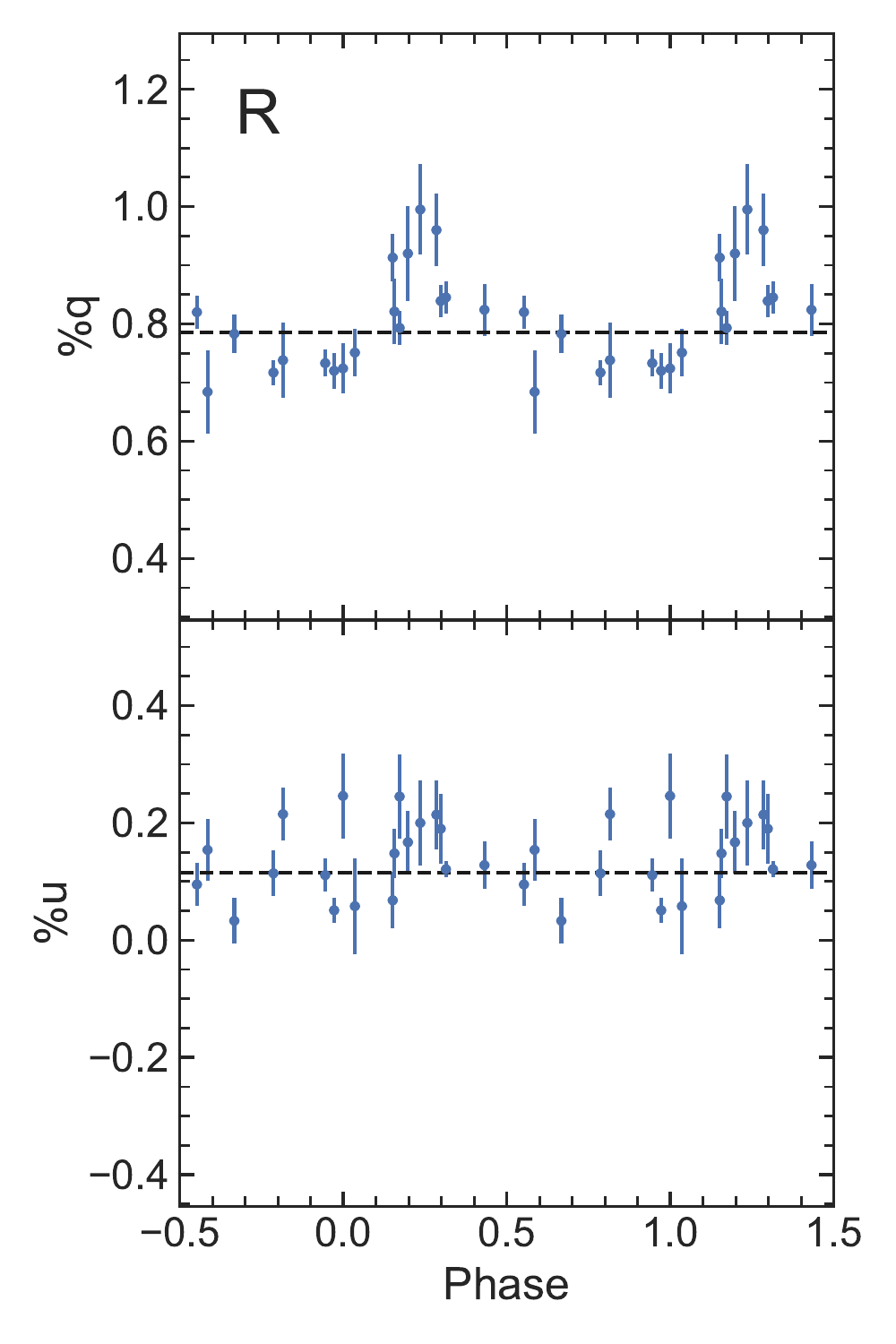}{0.33\textwidth}{(d)}
\fig{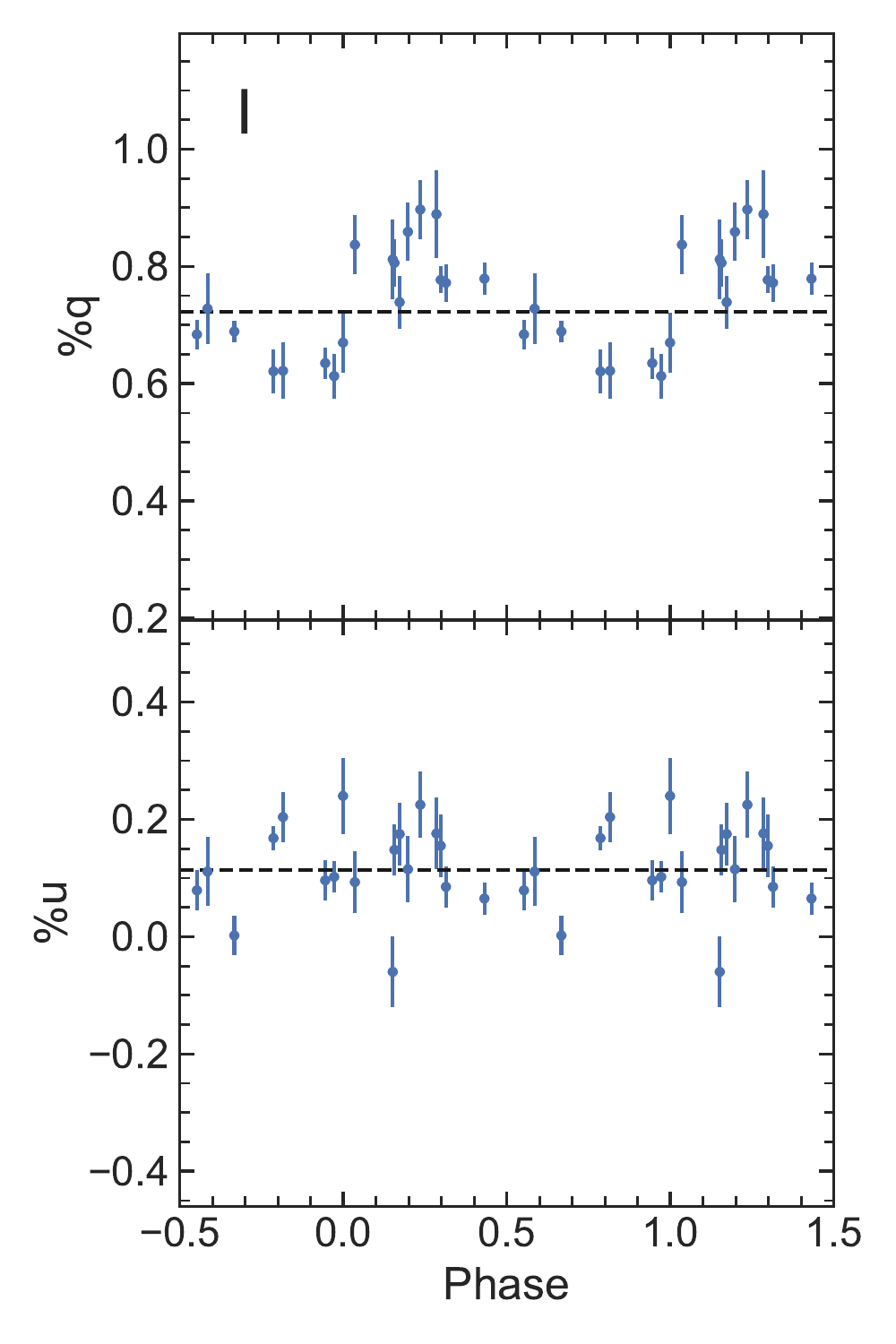}{0.33\textwidth}{(e)}
\fig{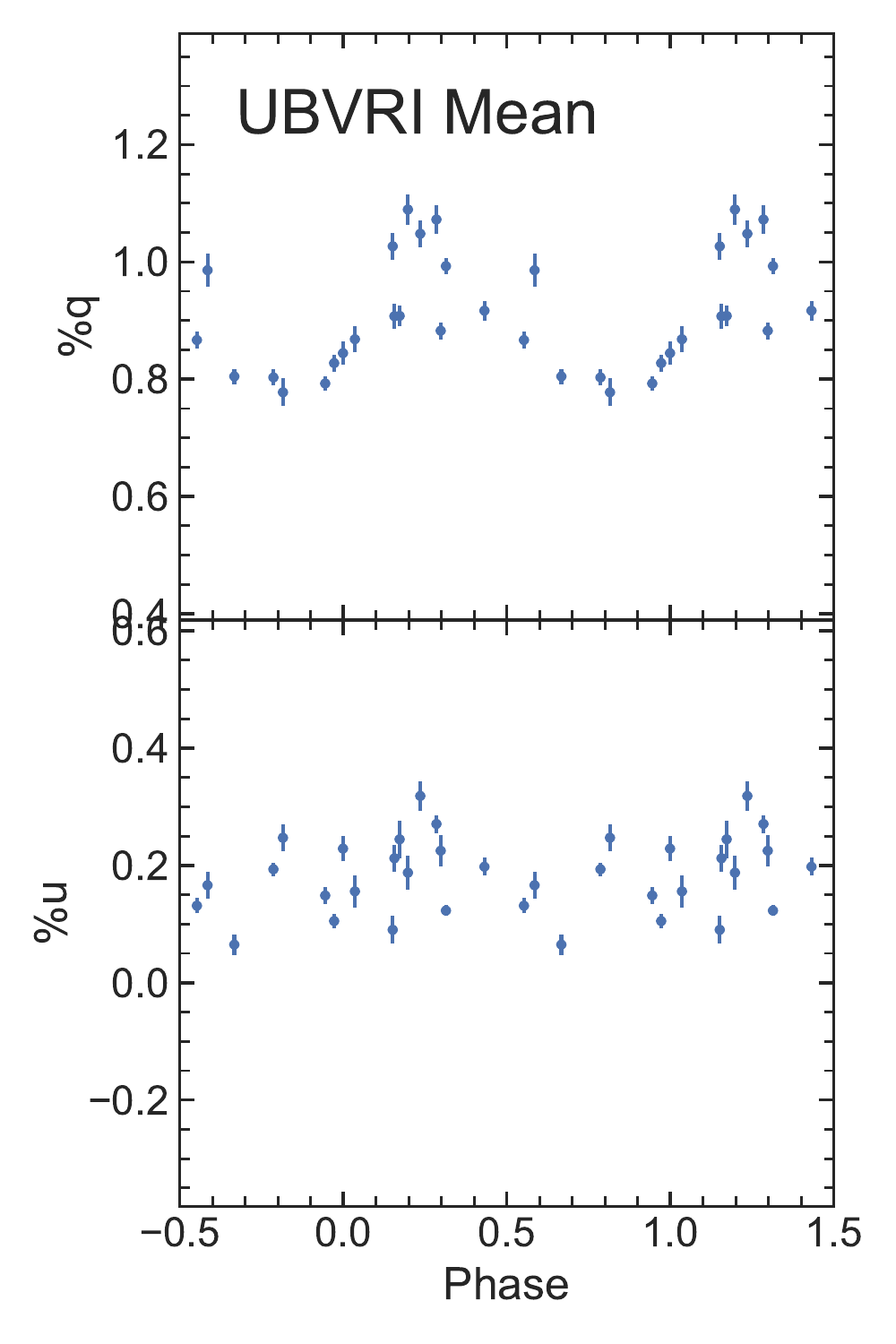}{0.33\textwidth}{(f)}}
\caption{$UBVRI$ filter data for WR 134 (Section~\ref{sec:binaryfit}\textbf{c}). Panels (a) through (e) correspond to $U$ through $I$ filters. The dashed line shows the weighted mean polarization value in each band. Panel (f) displays the uncertainty-weighted mean of the five filters. The data presented in this figure are available in the Appendix, Table~\ref{tab:wr134data}.}
\label{fig:wr134data}
\end{figure*}

\textbf{d) WR 6} The binary status of this object has been in dispute for many years. It has been proposed that its periodic variability can be explained by CIRs \citep[e.g.][]{moffat_wolf-rayet_2018, st-louis_polarization_2018}, or by the apsidal motion of a binary companion \citep[e.g.][]{schmutz_long_2019}. Given the uncertainty surrounding the nature of the object, and the limited number of data points in our sample, we simply take a per-band weighted mean of the $UBVRb$ data presented in \citet{moffat_polarization_1993}.

\textbf{e) WR 42, WR 79, WR 155} These systems are all binaries, and their systemic means were already published by \citet{moffat_polarization_1993} (WR 42, WR 79) and \citet{piirola_simultaneous_1988} (WR 155), produced using the model of \citet{brown_polarisation_1978}. We provide their values for reference purposes in Table~\ref{tab:snapshotdata}.

\textbf{f) WR 48, WR 113} These systems are binaries, but we observed them only twice each. Thus, it is not feasible to fit binary models to these data, so we took an uncertainty-weighted mean in each band instead of attempting to fit each observation separately.


\textbf{g) WR 16, WR 40, WR 103} These systems exhibit significant random polarization variation. As in \textbf{c)} and \textbf{d)}, we took an uncertainty-weighted mean in each band for each system.

\textbf{h) WR 22, WR 43, WR 71, WR 111} These systems showed no polarization variability greater than $2\sigma$ over multiple nights. We took an uncertainty-weighted mean in each band, even for the binary WR 22 and the pair of binaries in WR 43. 


\section{Simultaneous fit of interstellar and intrinsic polarization \label{sec:fitting}}

We next investigated the contribution of interstellar polarization to each of our targets. Using the mean polarization values we derived in Section~\ref{sec:binaryfit}, we followed \citet{moffat_polarization_1993} to fit $q$ and $u$ simultaneously for all objects with a modified Serkowski law:
\begin{multline}\label{eqn:qisp}
    q = q_{00} + P_\mathrm{IS,max}\cos{2\theta_\mathrm{IS}} \\
    \times\exp \,[-1.7\lambda_\mathrm{max}\ln^2(\lambda_\mathrm{max}/\lambda)]
\end{multline}
\begin{multline}\label{eqn:uisp}
    u = u_{00} + P_\mathrm{IS,max}\sin{2\theta_\mathrm{IS}} \\
    \times\exp \,[-1.7\lambda_\mathrm{max}\ln^2(\lambda_\mathrm{max}/\lambda)].
\end{multline}
\noindent In these equations, $q_{00}$ and $u_{00}$ represent constant polarization intrinsic to the system, which we expect to be independent of wavelength due to free-electron scattering in WR winds. Given the measurement uncertainties in our data, wavelength-dependent effects in this intrinsic polarization (due to dust scattering or absorption in the WR environment) are unlikely to be detectable. $P_\mathrm{IS,max}$ represents the peak interstellar polarization value and $\lambda_\mathrm{max}$ the wavelength at which this peak occurs. These equations follow the prescription of \citet{wilking_wavelength_1980}, in which the constant $K$ in the classic Serkowski law \citep{serkowski_wavelength_1975} is replaced by $1.7\lambda_{max}$. As in \citet{moffat_polarization_1993}, we allow the position angle of the ISP to vary inversely with wavelength: $\theta_\mathrm{IS} = \theta_\mathrm{0} + k/\lambda$ . 

As in Section~\ref{sec:binaryfit}, we carried out the fits using \textsc{lmfit}, beginning with the least-squares Levenberg-Marquadt method, then using the Markov-Chain Monte Carlo fitting module \textsc{emcee} \citep{foreman-mackey_emcee:_2013} as part of the \textsc{lmfit} module to refine the fits. We constrained the variable $\lambda_\mathrm{max}$ to lie in the range 0.35--1.0 \micron\ . 
We chose initial parameter values from the data: $P_\mathrm{IS,max}$ began as the maximum total polarization across all bands, $\theta_\mathrm{0}$ began as the average position angle across all bands, and $\lambda_\mathrm{max}$ began as the central wavelength of the filter with the maximum total polarization value. We omitted data from La Silla in the additional Str\"omgren $b$ filter 
because this filter can be potentially strongly affected by depolarization in the $\lambda$4650 line region (comprised of \ion{C}{3} $\lambda$4650 + \ion{C}{4} $\lambda$4658 + \ion{He}{2} $\lambda$4686 in WC stars, or \ion{He}{2} $\lambda$4686 + \ion{N}{5} $\lambda$4601/4604/4619 + \ion{N}{3} $\lambda$4634-4642 in WN stars). While other filters may also be affected by line depolarization, the $\lambda$4650  region contains the strongest lines in the WR optical spectrum, and the Str\"omgren $b$ filter is significantly narrower in wavelength than any of the Johnson filters. The \textit{UBVRI} data are therefore much less susceptible to line depolarization effects than the Str\"omgren $b$ data. Thus, we neglect any line contributions to our broadband polarization results.

\begin{figure*}
    \centering
    \includegraphics[width=0.95\textwidth]{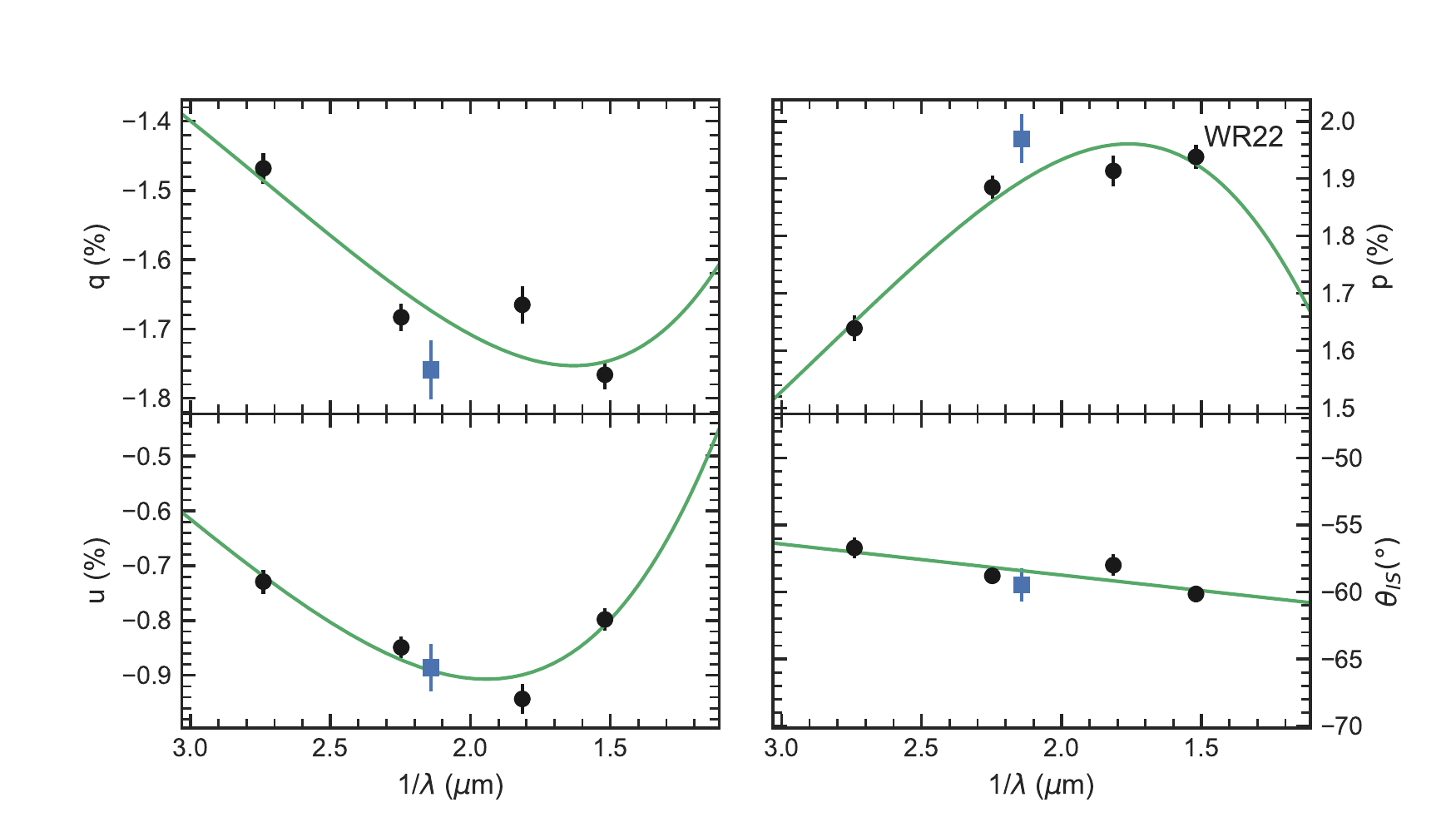}
    \caption{$UBVRI$ polarization data for WR 22 (black points) fitted with equations \ref{eqn:qisp} and \ref{eqn:uisp} (green curves). Str\"omgren $b$ filter polarization data are shown as blue points. The position angle points have been shifted by subtracting the fitted $q_{00}$ and $u_{00}$ values from the original data (Section~\ref{sec:fitting}).}
    \label{fig:wr22_example}
\end{figure*}

Figure \ref{fig:wr22_example} shows an example fit to the data for WR 22, using equations \ref{eqn:qisp} and \ref{eqn:uisp}. The left panel shows the data that were fitted, while the right shows the same data and fit transformed to the usual $p$ and $\theta$ space of the Serkowski law. In this case, the parameter $k$ has $>3\sigma$ significance (i.e. $|k|>3\sigma_k$. In order to depict the wavelength dependence of  $\theta_{IS}$, we subtracted the fitted $q_{00}$ and $u_{00}$ parameters from the data and recalculated the position angle displayed in the figure. 


After the first round of fits, we checked whether the fitted values for the intrinsic components $q_{00}$ and $u_{00}$ were significant, taking significant values to be at least 2$\sigma$ above the estimated fit uncertainties, derived from the MCMC posterior probability distribution for each parameter. If the result for a given star was not significant for those parameters, we repeated the fit using the equations

\begin{equation}\label{eqn:qispshort}
    q = P_\mathrm{IS,max}\cos{2\theta_\mathrm{IS}} \exp \,[-1.7\lambda_\mathrm{max}\ln^2(\lambda_\mathrm{max}/\lambda)]
\end{equation}
\begin{equation}\label{eqn:uispshort}
    u = P_\mathrm{IS,max}\sin{2\theta_\mathrm{IS}} \exp \,[-1.7\lambda_\mathrm{max}\ln^2(\lambda_\mathrm{max}/\lambda)].
\end{equation}

\begin{figure*}
    \centering
    \includegraphics[width=0.95\textwidth]{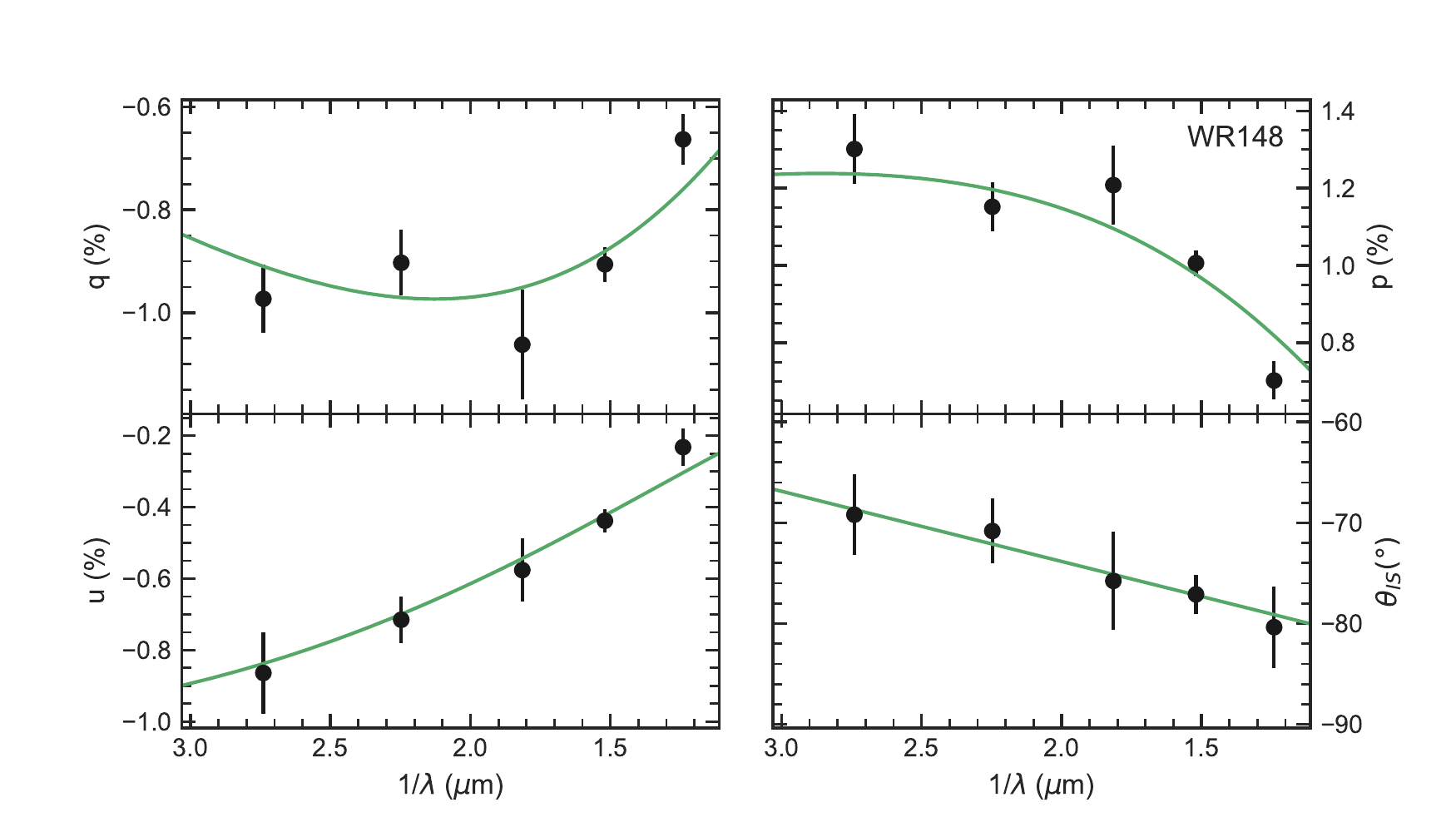}
    \caption{$UBVRI$ polarization data for WR 148 (black points), fitted with equations \ref{eqn:qispshort} and \ref{eqn:uispshort} (green curves).}
    \label{fig:wr148_example}
\end{figure*}

\noindent This was done to ensure accurate ISP estimates in cases where the uncertainties on $q_{00}$ and $u_{00}$ were large. In those cases, the uncertainty in other parameters grew larger and reduced the significance of the $k$ parameter result. Figure \ref{fig:wr148_example} shows an example fit to the data for WR 148 using equations \ref{eqn:qispshort} and \ref{eqn:uispshort}. 

\begin{figure*}
    \plotone{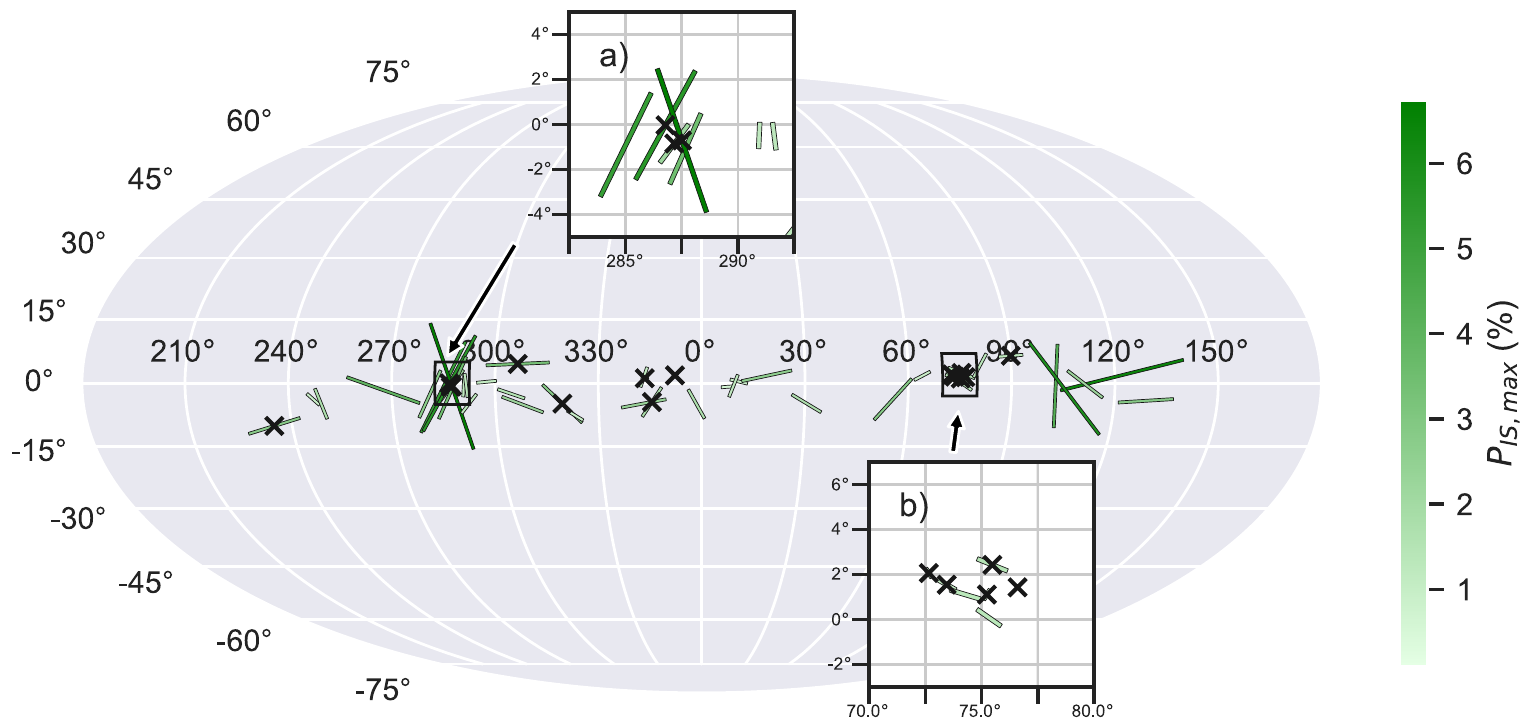}
    \caption{Map of our WR sample in Galactic coordinates, depicting our fitted polarization and position angle values for each star (Section~\ref{sec:fitting}; Table~\ref{tab:snapshot_isp}). The length of the bars is proportional to $P_\mathrm{IS,max}$. The angle of each bar represents $\theta_0$, measured counterclockwise from the horizontal 0$\degr$ line. Black crosses represent stars with $k/\sigma_k > 3$. Inset a) shows the region including WR 22, 23, and 25. Inset b) shows the region including WR 133, 134, 136, 138, and 139. We discuss these regions in Section~\ref{sec:wavdiscussion}.}
    \label{fig:isp_skymap}
\end{figure*}

We adopted final parameter values from the maximum likelihood estimates provided by \textsc{emcee} for all objects. We calculated 1$\sigma$ error estimates from the 1$\sigma$ Gaussian percentile of each parameter posterior probability distribution produced by \textsc{emcee}. We present the fitting results in Table~\ref{tab:snapshot_isp}, with values derived from equations \ref{eqn:qispshort} and \ref{eqn:uispshort} indicated in boldface. Figures displaying fits for all systems are available as online material (see Figure Set~\ref{fig:figureset}). We plot the results on a map of the sky in Galactic coordinates in Figure~\ref{fig:isp_skymap}.

\begin{longrotatetable}
\begin{deluxetable*}{lccccDDDDDDDD}
\tablecaption{Results from our fits of interstellar + constant intrinsic polarization contributions to each of our targets (Section~\ref{sec:fitting}.)
\label{tab:snapshot_isp}}
\tablehead{WR  & $q_{00}$ (\%)   & $\sigma_{q_{00}}$ (\%) & $u_{00}$  (\%)  & $\sigma_{u_{00}}$ (\%) & \multicolumn2c{$P_\mathrm{IS,max}$ (\%)} & \multicolumn2c{$\sigma_{P_\mathrm{IS,max}}$ (\%)} & \multicolumn2c{$\lambda_{max}$ (\micron)} & \multicolumn2c{$\sigma_{\lambda_{max}}$ (\micron)} & \multicolumn2c{$\theta_{0}~(\degr)$} & \multicolumn2c{$\sigma_{\theta_{0}}~(\degr)$} & \multicolumn2c{$k~(\degr\micron)$}      & \multicolumn2c{$\sigma_k~(\degr\micron)$}}
\startdata
\decimals
1\tablenotemark{a}   & \nodata  & \nodata & \nodata  & \nodata & 6.420  & 0.034 & 0.491 & 0.007 & -83.0 & 0.7  & -0.3  & 0.4  \\
3 \tablenotemark{a}  & \nodata  & \nodata & \nodata  & \nodata & 2.806  & 0.070 & 0.504 & 0.028 & -88.2 & 2.2  & 1.0   & 1.2  \\
6   & -2.220\tablenotemark{b}  & 0.143 & -0.782\tablenotemark{b}  & 0.139 & 2.721 & 0.147 & 0.521 & 0.004 & -57.2 & 4.1 & -3.2 & 1.2 \\
8\tablenotemark{a}   & \nodata  & \nodata & \nodata  & \nodata & 0.860  & 0.023 & 0.499 & 0.035 & -20.9 & 3.5  & -1.9  & 1.7  \\
9\tablenotemark{a}   & \nodata  & \nodata & \nodata  & \nodata & 1.657  & 0.214 & 0.600 & 0.148 & -34.0 & 16.3 & -6.1 & 8.8 \\
14  & -1.151\tablenotemark{b} & 0.281 & 0.129  & 0.267 & 3.918  & 0.289 & 0.539 & 0.004 & -9.8  & 1.8  & -0.1  & 0.2  \\
16  & 0.293  & 0.426 & 0.957  & 0.468 & 2.632  & 0.500 & 0.548 & 0.023 & -57.2 & 4.1  & -3.2  & 1.2  \\
21  & 0.878  & 0.512 & 3.002\tablenotemark{b}  & 0.554 & 5.132  & 0.579 & 0.529 & 0.006 & -58.0 & 2.8  & -0.5  & 0.3  \\
22  & -0.728 & 0.325 & 0.979  & 0.354 & 2.135  & 0.371 & 0.538 & 0.015 & -63.4 & 4.2  & 2.3\tablenotemark{b}   & 0.7  \\
23  & 0.000  & 0.350 & 1.772\tablenotemark{b}  & 0.432 & 5.512  & 0.433 & 0.531 & 0.004 & -59.4 & 1.9  & -1.1\tablenotemark{b}  & 0.2 \\
24  & -0.211 & 0.480 & 1.305  & 0.534 & 3.433  & 0.549 & 0.535 & 0.012 & -56.9 & 3.6  & 0.8   & 0.6  \\
25\tablenotemark{a}  & \nodata  & \nodata & \nodata  & \nodata & 6.741  & 0.042 & 0.631 & 0.006 & -35.5 & 0.6  & -4.9\tablenotemark{b}  & 0.3  \\
40\tablenotemark{a}  & \nodata  & \nodata & \nodata  & \nodata & 1.234  & 0.018 & 0.585 & 0.015 & -63.9 & 1.7  & 0.2   & 0.8 \\
42  & -0.409\tablenotemark{b} & 0.124 & 0.568\tablenotemark{b}  & 0.127 & 1.177 & 0.131 & 0.568 & 0.009 & -46.3  & 2.5  & -0.4  & 0.5  \\
43\tablenotemark{a}  & \nodata  & \nodata & \nodata  & \nodata & 1.233  & 0.023 & 0.616 & 0.017 & -41.4 & 2.0  & -2.9  & 1.0  \\
46\tablenotemark{a}  & \nodata  & \nodata & \nodata  & \nodata & 1.006  & 0.025 & 0.525 & 0.026 & -87.2 & 2.7  & 0.1   & 1.2  \\
48\tablenotemark{a}  & \nodata  & \nodata & \nodata  & \nodata & 1.446  & 0.014 & 0.551 & 0.013 & 80.6  & 1.3  & -0.2  & 0.7  \\
52\tablenotemark{a}  & \nodata  & \nodata & \nodata  & \nodata & 3.208  & 0.018 & 0.579 & 0.006 & 91.2  & 0.6  & -2.4\tablenotemark{b} & 0.3  \\
57\tablenotemark{a}  & \nodata  & \nodata & \nodata  & \nodata & 2.233  & 0.031 & 0.579 & 0.012 & 79.7  & 1.4  & -1.4  & 0.7  \\
69\tablenotemark{a}  & \nodata  & \nodata & \nodata  & \nodata & 2.775  & 0.020 & 0.540 & 0.008 & 68.1  & 0.9  & -4.1\tablenotemark{b}  & 0.4  \\
71  & 0.171  & 0.336 & 0.592  & 0.254 & 0.939  & 0.373 & 0.586 & 0.110 & 74.3  & 9.2  & -0.2 & 6.7  \\
78\tablenotemark{a}  & \nodata  & \nodata & \nodata  & \nodata & 1.063  & 0.017 & 0.675 & 0.016 & 35.8  & 1.8  & 0.9   & 1.0  \\
79\tablenotemark{a} & \nodata  & \nodata & \nodata  & \nodata & 0.376 & 0.007 & 0.595 & 0.015 & -81.5  & 1.8  & 4.7\tablenotemark{b} & 0.8  \\
86\tablenotemark{a}  & \nodata  & \nodata & \nodata  & \nodata & 0.370  & 0.015 & 0.654 & 0.054 & 108.9 & 5.7  & -26.2\tablenotemark{b} & 3.3  \\
90  & -0.918 & 0.647 & 1.993\tablenotemark{b} & 0.353 & 2.296  & 0.667 & 0.520 & 0.026 & 5.3   & 8.5  & -4.1  & 2.2  \\
92\tablenotemark{a} & \nodata  & \nodata & \nodata  & \nodata & 1.770  & 0.022 & 0.560 & 0.013 & 28.4  & 1.4  & -3.0\tablenotemark{b} & 0.7  \\
103 & -0.889\tablenotemark{b} & 0.232 & 0.561  & 0.257 & 1.682  & 0.268 & 0.518 & 0.008 & -30.3 & 4.1  & -0.2  & 0.4  \\
108 & 1.089\tablenotemark{b}  & 0.213 & 0.759\tablenotemark{b}  & 0.155 & 0.585  & 0.255 & 0.701 & 0.190 & 2.0   & 49.2 & -9.8  & 31.5 \\
110\tablenotemark{a} & \nodata  & \nodata & \nodata  & \nodata & 0.904  & 0.022 & 0.551 & 0.040 & -6.7  & 3.7  & -1.1  & 2.0  \\
111 & -0.111 & 0.212 & 0.940  & 0.376 & 1.208  & 0.392 & 0.544 & 0.036 & -55.8 & 8.6  & 4.1   & 3.3  \\
113\tablenotemark{a} & \nodata  & \nodata & \nodata  & \nodata & 2.592  & 0.020 & 0.522 & 0.011 & -84.0 & 1.1  & -1.4  & 0.6  \\
123\tablenotemark{a} & \nodata  & \nodata & \nodata  & \nodata & 1.729  & 0.032 & 0.585 & 0.021 & 74.4  & 2.3  & 3.0   & 1.2  \\
127\tablenotemark{a} & \nodata  & \nodata & \nodata  & \nodata & 0.916  & 0.034 & 0.528 & 0.042 & 13.9  & 4.3 & 4.5   & 2.2  \\
128 & -0.621 & 0.440 & -1.696\tablenotemark{b} & 0.470 & 2.811  & 0.494 & 0.496 & 0.022 & 23.8  & 6.5  & 0.0 & 0.9  \\
133\tablenotemark{a} & \nodata & \nodata & \nodata  & \nodata & 0.491  & 0.009 & 0.400 & 0.020 & -27.7 & 1.5  & 4.4\tablenotemark{b} & 0.7  \\
134 & 0.112  & 0.065 & 0.307\tablenotemark{b}  & 0.067 & 0.942  & 0.060 & 0.350 & 0.004 & -14.7 & 5.0  & 5.0\tablenotemark{b} & 1.1  \\
135\tablenotemark{a} & \nodata  & \nodata & \nodata  & \nodata & 0.162  & 0.032 & 0.350 & 0.133 & 19.3  & 24.1 & -3.5  & 12.7 \\
136\tablenotemark{a} & \nodata  & \nodata & \nodata  & \nodata & 1.454  & 0.018 & 0.490 & 0.017 & -10.9 & 1.5  & 4.0\tablenotemark{b}   & 0.8  \\
137\tablenotemark{a} & \nodata  & \nodata & \nodata  & \nodata & 1.293  & 0.027 & 0.498 & 0.026 & -8.3  & 2.3  & -0.7  & 1.1 \\
138\tablenotemark{a} & \nodata  & \nodata & \nodata  & \nodata & 0.538  & 0.021 & 0.522 & 0.050 & -62.1 & 4.8  & -9.6\tablenotemark{b} & 2.5  \\
139 & 0.022  & 0.041 & -0.267\tablenotemark{b} & 0.089 & 0.098  & 0.083 & 0.970 & 0.065 & 83.8 & 19.6 & -47.0\tablenotemark{b}  & 6.0  \\
140\tablenotemark{a} & \nodata  & \nodata & \nodata  & \nodata & 1.369  & 0.018 & 0.519 & 0.015 & 31.0  & 1.1  & 0.0 & 0.6  \\
141\tablenotemark{a} & \nodata  & \nodata & \nodata  & \nodata & 1.309  & 0.016 & 0.526 & 0.015 & 72.6  & 1.3  & 1.3  & 0.7  \\
148\tablenotemark{a} & \nodata  & \nodata & \nodata  & \nodata & 1.238  & 0.036 & 0.350 & 0.018 & -87.8 & 2.6  & 7.0\tablenotemark{b} & 1.4  \\
153\tablenotemark{a} & \nodata  & \nodata & \nodata  & \nodata & 4.242  & 0.017 & 0.541 & 0.005 & 43.7  & 0.8  & 1.2\tablenotemark{b} & 0.4  \\
155 & 0.412\tablenotemark{b} & 0.119 & -0.575\tablenotemark{b} & 0.120 & 5.842  & 0.126 & 0.515 & 0.002 & 63.6  & 0.7  & 0.1 & 0.1 \\
157\tablenotemark{a} & \nodata  & \nodata & \nodata  & \nodata & 2.250  & 0.045 & 0.465 & 0.020 & 71.2  & 1.6  & 1.1   & 0.9 
\enddata
\tablenotetext{a}{Results fitted using equations \ref{eqn:qispshort} and \ref{eqn:uispshort}.}
\tablenotetext{b}{Results with 3$\sigma$ significance.}
\tablecomments{Targets were fit using equations~\ref{eqn:qisp} and~\ref{eqn:uisp} unless marked.} 
\end{deluxetable*}
\end{longrotatetable}

\section{Discussion}

\subsection{Intrinsic polarization\label{sec:intrinsicpoldiscussion}}

The intrinsic continuum polarization ($q_{00}$, $u_{00}$ values) we detect in some of our targets could originate from the asymmetric illumination of a spherically symmetric free electron distribution or from a symmetric illumination of an asymmetric distribution (or both). For single WR stars the polarization is likely caused by light from the WR star scattering in an asymmetric wind \citep[e.g.,][]{harries_spectropolarimetric_1998,st-louis_revealing_2013}. In close WR binary systems all the above-mentioned effects can take place. In the case of the systems for which we estimated the binary polarization variations using the \citet{brown_polarisation_1978} model (\S~\ref{sec:binaryfit}), the remaining intrinsic polarization contributions could still be due to optically thick scattering or a finite stellar disk, which are not taken into account in that model. However, an examination of the results of \citet{vink17wolf} for the SMC and LMC indicates that binaries are no more likely that single stars to posses intrinsic continuum polarization. For the Galaxy, the results of \citet{harries_spectropolarimetric_1998} lead to similar conclusions. This seems to suggest that although a binary effect is expected in close WR + O systems \citep[e.g.][]{st-louis_polarization_1988}, the probability of detecting it in a single snapshot observation is low. Therefore, in binaries for which we obtained only a single measurement of intrinsic continuum polarization at an arbitrary phase, or could not characterize the time-dependent polarization variations for other reasons, the intrinsic polarization may still include these binary effects. In these cases we cannot constrain the polarization source without additional phase-dependent observations.


Based on the fits described in Section~\ref{sec:fitting}, 12 stars in our sample showed intrinsic polarization  above the 3$\sigma$ level. However, WR 108 and WR 139 are outliers  in this group because they do not have clearly defined values of $P_\mathrm{IS,max}$ within the observed $UBVR$ wavelength range. This means that the ISP toward them is also poorly defined, as shown by the large uncertainty on their polarization position angles (Table~\ref{tab:snapshot_isp}). As a result, their intrinsic polarization values are also poorly defined, regardless of the formal uncertainties, and we do not consider that we have detected significant intrinsic polarization for these stars.

Of the remaining 10 targets with intrinsic polarization, WR 21, WR 42, and WR 155 are known short-period binaries.
In the case of WR 42, a short-period WC7 + O7V binary, 
we used the systemic mean polarization from binary fits made using the model of \citet{brown_polarisation_1978} (Section \ref{sec:binaryfit}\textbf{e}).
Thus the additional intrinsic polarization in WR 42 must be due to a wind asymmetry that is not incorporated into this model. Such an asymmetry could be caused by the binary interactions modeled by \citet{hill_modelling_2000} or rapid rotation of the WR star, and warrants further study with time-dependent polarimetry. 

WR 155 is an extremely short-period WN6o + O9II-Ib system, for which we also used systemic mean polarization values from previous fits (\citealt{piirola_simultaneous_1988}; Section \ref{sec:binaryfit}\textbf{e}). This system undergoes sporadic periods of Roche lobe overflow, transferring mass between the O and WR stars \citep{koenigsberger_does_2017}. It is therefore likely that the intrinsic polarization is caused by asymmetric wind structures produced due to these interactions between the stars.

We obtained only one  snapshot observation of WR 21. The intrinsic Stokes $u$ of $3.002\%\pm0.554\%$ resulting from our fit should be treated with caution since it is much larger than any of our other measurements; further investigation is needed to check this result. Such a high polarization is not unprecedented, however; \citet{villarsbaffi06extreme} found an intrinsic level of $3-4\%$ in the short-period WR + O binary WR 151 (CX Cep).  
For the range of inclination angles derived for WR 21 by \citealt{lamontagne_photometric_1996} (48--62$\degr$), we calculate that the model of \citet{brown_polarisation_1978} produces a maximum polarization of $P=0.1-0.4\%$ (regardless of the value of $\Omega$). This is consistent with the amplitude of modulation we have found in additional unpublished data. If we take WR 21's large intrinsic polarization at face value, then, it is very unlikely to be due to binary effects alone. We thus hypothesize that WR 21 contains an asymmetric WR wind, which may be as extremely flattened as that of WR 151 \citep{villarsbaffi06extreme}. Our Str\"omgren $b$ filter results in \S~\ref{sec:stromgren} show no significant line depolarization for this object, but this does not necessarily imply a spherical wind \citep{stevance_probing_2018}. It is also important to note that our single observation does not preclude the existence of a transient, high-density clump. Further observations of WR 21 at different orbital phases would help clarify the situation. 

Of the 7 other probably single stars with significant intrinsic polarization, WR 134 has been found to harbour complex wind structures \citep{aldoretta_extensive_2016}, which likely give rise to the observed intrinsic polarization. WR 128 is a WN4(h) type with small-scale spectral variability that may indicate inhomogeneities or clumps in the wind which could also cause a polarization signal \citep{st-louis_systematic_2009}. 

WR 6 is a WN4b star with a possible companion \citep[e.g.][]{schmutz_long_2019} or CIRs \citep[e.g.][]{moffat_wolf-rayet_2018, st-louis_polarization_2018}. \citet{harries_interstellar_1999} measured the ISP using a different method from ours, and found a very different result of $P_{max} = 0.47\pm0.02\%$ at $\theta_0 = 164\pm2\degr$. This agreed with \citet{robert_photometry_1992} and \citet{schulte-ladbeck_wind_1991}. However, they did not simultaneously fit the intrinsic polarization, nor did they include a wavelength-dependent position angle. On the other hand, as we discuss in Section~\ref{sec:stromgren}, our $B$-band $u$ measurement was strongly affected by the depolarization of emission lines in the system, and this may affect our fits. This complex system needs more spectropolarimetric observations to resolve its nature and measure the true value of its ISP.

The remaining 4 stars are all late-type WC types. WR 14, WR 23, and WR 103 display a relatively high level of small-scale spectral variability characteristic of strong clumping in their winds \citep{michaux_origin_2014}, and this is most likely the cause of the (variable) intrinsic polarization. This variability was also detected in polarimetry by \citet{drissen_polarization_1992} in the case of WR 14. WR 90 shows a residual in the $b$ filter measurement; we discuss this object in more detail in Section~\ref{sec:stromgren} below.

Table~\ref{tab:limits} lists our findings for the intrinsic polarization (assumed constant with wavelength)
of all objects in our sample. In cases where $|q_{00}| > 2\sigma_{q_{00}}$ or $|u_{00}| > 2\sigma_{u_{00}}$, we display our fitted quantities (uncertainties on these quantities are shown in Table~\ref{tab:snapshot_isp}). Otherwise, we quote upper absolute limits based on the $1\sigma$ observational uncertainties, or fit uncertainties in the case of stars with multiple observations. These were calculated as a mean over $UBVR$ uncertainties (and $I$ when available;  Table~\ref{tab:snapshotdata}) in each of $q$ and $u$. The band-to-band uncertainties are consistent at the $\sim0.06\%$ level for $U$ and $V$, and the $\sim0.04\%$ level for $B$, $R$, and $I$. These values can be used to guide the required precision of future polarization observations of these systems.

\startlongtable
\begin{deluxetable}{lCC}
\label{tab:limits}
\tablecaption{Intrinsic polarization values and limits for the WR stars in our sample. 
}
\tablehead{\colhead{WR}  & \colhead{\hspace{1.25cm}$q_{00}$~(\%)}\hspace{1.25cm} & \colhead{\hspace{1.1cm}$u_{00}$~(\%)}\hspace{1.25cm}}
\startdata
1   & <0.07 & <0.06 \\
3   & <0.15 & <0.13 \\
6   & -2.220 & -0.782 \\
8   & <0.04 & <0.04 \\
9   & <0.36 & <0.36 \\
14  & -1.151 & <0.02 \\
16  & <0.03 & \phm{-}0.957 \\
21  & <0.04 & \phm{-}3.002 \\
22  & -0.728 & \phm{-}0.979 \\
23  & <0.03 & \phm{-}1.772\\
24  & <0.04 & \phm{-}1.305 \\
25  & <0.06 & <0.06 \\
40  & <0.03 & <0.03 \\
42  & -0.409 & \phm{-}0.568 \\
43  & <0.03 & <0.03 \\
46  & <0.04 & <0.04 \\
48  & <0.03 & <0.03 \\
52  & <0.03 & <0.03 \\
57  & <0.05 & <0.05 \\
69  & <0.03 & <0.03 \\
71  & <0.03 & \phm{-}0.592 \\
78  & <0.03 & <0.03 \\
79  & <0.01 & <0.01 \\
86  & <0.04 & <0.04 \\
90  & <0.04 & \phm{-}1.993 \\
92  & <0.0 & <0.04 \\
103 & -0.889 & \phm{-}0.561 \\
108 & <0.07 & <0.07 \\
110 & <0.05 & <0.05 \\
111 & <0.02 & \phm{-}0.940 \\
113 & <0.04 & <0.04 \\
123 & <0.06 & <0.06 \\
127 & <0.07 & <0.08 \\
128 & <0.07 & -1.696 \\
133 & <0.01 & <0.01 \\
134 & <0.01 & \phm{-}0.307 \\
135 & <0.08 & <0.07 \\
136 & <0.04 & <0.04 \\
137 & <0.06 & <0.06 \\
138 & <0.05 & <0.05 \\
139 & <0.04 & -0.267 \\
140 & <0.03 & <0.04 \\
141 & <0.03 & <0.03 \\
148 & <0.06 & <0.07 \\
153 & <0.07 & <0.04 \\
155 & \phm{-}0.412 & \phm{-}0.575 \\
157 & <0.08 & <0.09
\enddata
\tablecomments{We present fitted  results when they are at least 2$\times$ greater than the $\sigma_{q_{00}}$ or $\sigma_{u_{00}}$ uncertainty displayed in Table~\ref{tab:snapshot_isp}. Otherwise, we present upper limits computed using the mean 1$\sigma$ broadband polarization uncertainties from our observational weighted means or systemic mean calculations. Band-to-band uncertainties are consistent at the $\sim0.06\%$ level for $U$ and $V$, and the $\sim0.04\%$ level for $B$, $R$, and $I$.}
\end{deluxetable}

\subsection{The wavelength dependence of the ISP position angle\label{sec:wavdiscussion}}

\citet{dolan_wavelength_1986} studied the optical wavelength dependence of linear polarization in a number of strongly polarized stars. For 9 of 11 such stars they found a significant deviation from a constant polarization angle and fitted their data with a wavelength ($\lambda$) dependent function. These curves can be better linearized in most cases by using $1/\lambda$ as the independent variable, which we adopted in our fits in search of a significant slope in each of our targets (Section~\ref{sec:fitting}). \citet{dolan_wavelength_1986} concluded that most of the nonzero slopes they derived could be intrinsic to the star, although they could not eliminate the presence of multiple dust clouds along the line of sight, each with different grain alignments. 
However, allowing for this effect is important, both to achieve the best possible fits to the ISP Serkowski law and to account for the possible presence of an intrinsic polarization component. 

\begin{figure}
    \centering
    \includegraphics[width=0.45\textwidth]{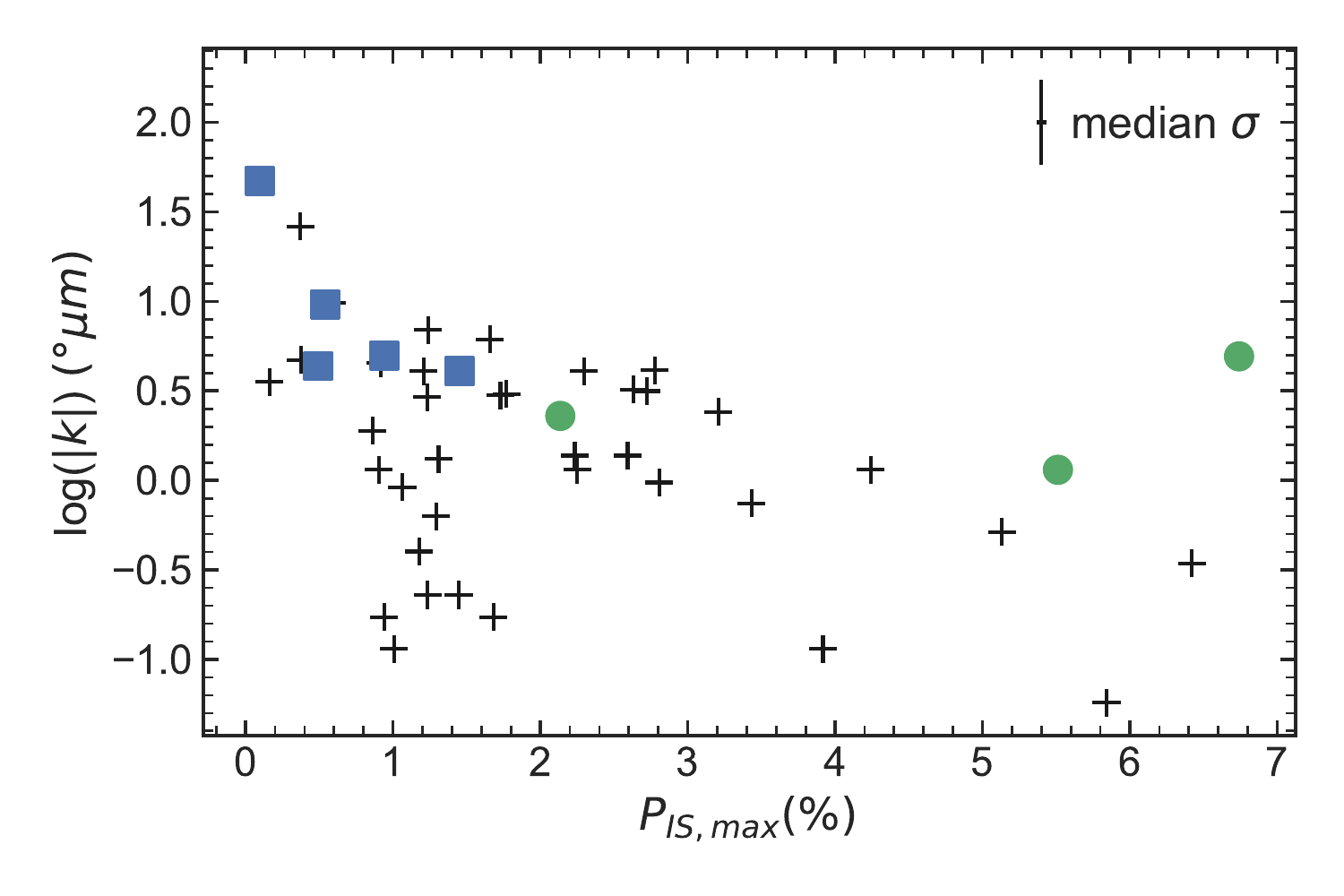}
    \caption{Interstellar position angle rotation coefficients (on a log scale) versus interstellar $P_\mathrm{IS,max}$ values for all stars in our sample. For clarity, we do not plot error bars on each point; median uncertainties for each quantity are represented by the cross-hairs in the upper right of the plot. Green circles correspond to the clustered stars in Fig.~\ref{fig:isp_skymap}, inset a). Blue squares correspond to the clustered stars in Fig.~\ref{fig:isp_skymap}, inset b).}
    \label{fig:pmax_vs_k}
\end{figure}

The recent ISP survey by \citet{bagnulo_large_2017} 
found that stars with strong wavelength dependence in the ISP position angles (large $|k|$) tend to have low interstellar polarization overall (small $P_\mathrm{IS,max}$). Our data confirm this trend, as shown in Figure~\ref{fig:pmax_vs_k}, which displays a weak inverse relationship between $P_\mathrm{IS,max}$ and $|k|$. 
This likely reflects the fact that as $P$ values become small (declining redward from typical  $\lambda_\mathrm{max}$ values of $\sim540$ nm), $\theta$ becomes less well defined, giving rise to apparent rotations with wavelength.

\begin{figure}
    \centering
    \includegraphics[width=0.45\textwidth]{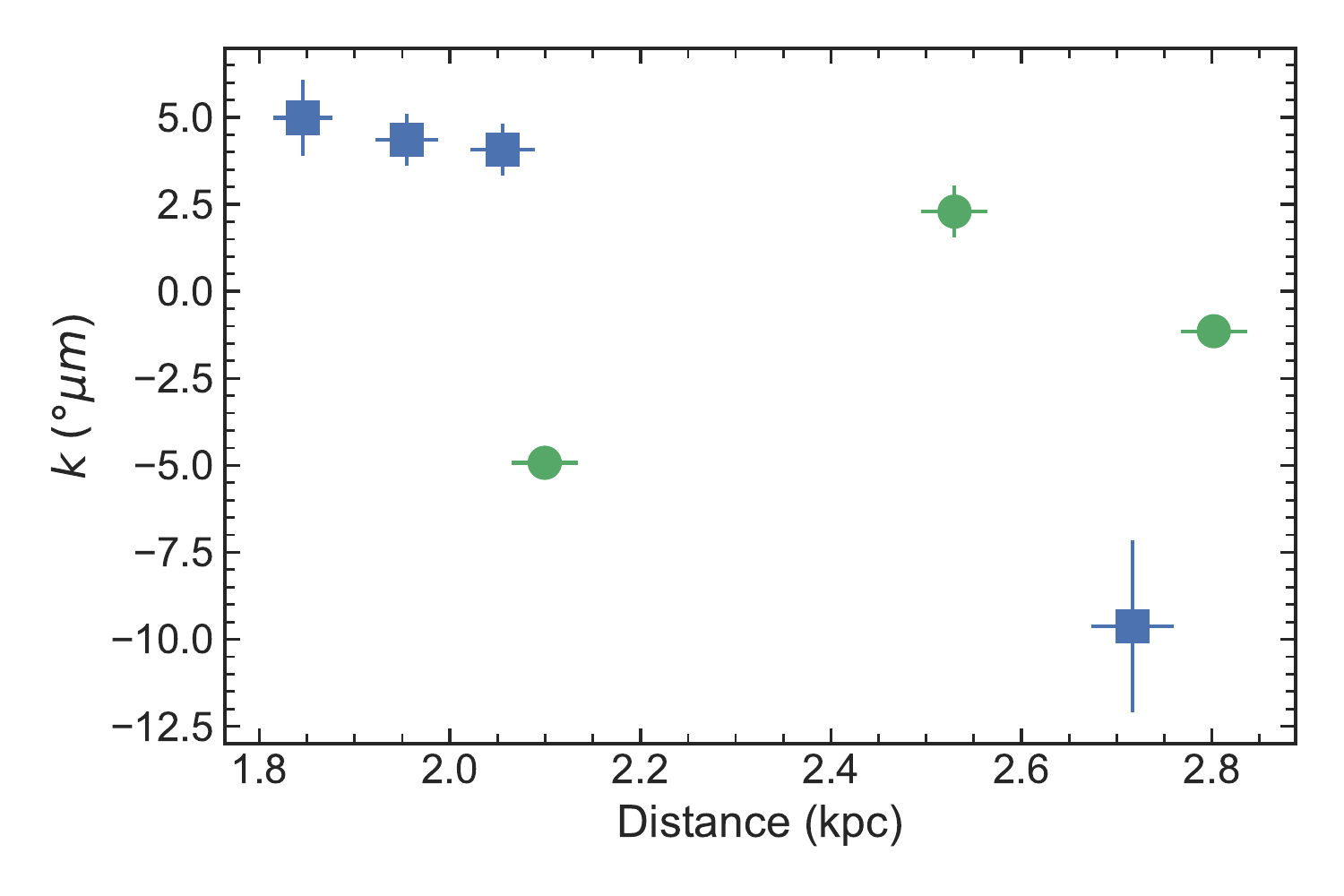}
    \caption{Interstellar position angle rotation coefficients versus Gaia DR2 distances  for the two star clusters displayed in the insets in Fig.~\ref{fig:isp_skymap}. Green circles correspond to systems in inset a) (WR 22, 23, 25). Blue squares correspond to systems in inset b) (WR 133, 134, 136, 138). Uncertainties in distance are derived from the Gaia data. WR 139 has been omitted due to its poorly defined $P_\mathrm{IS,max}$ (Section~\ref{sec:intrinsicpoldiscussion}).}
    \label{fig:distance_vs_k}
\end{figure}

Fifteen stars in our sample have ISP position angles ($\theta_\mathrm{IS}$) with significant wavelength dependence ($k>3\sigma_k$). In cases with low $P_\mathrm{IS,max}$, this significance may simply be due to the relation shown in Fig.~\ref{fig:pmax_vs_k}. However, within this subsample, two groups of stars stand out because they are clustered on the sky (as shown in the insets to Fig.~\ref{fig:isp_skymap}). WR 22, WR 23, and WR 25 lie within $\sim1\degr$ 
and have distances in the range 2.1--2.8 kpc \citep{gaia_collaboration_gaia_2018}. WR 133, WR 134, WR 136, and WR 138 lie within $\sim3\degr$ 
and have distances in the range 1.9--2.7 kpc. This clustering of stars with significant $\theta_\mathrm{IS}$ wavelength dependence supports the idea that this effect is due to scattering in multiple dust clouds along the line of sight. Figure~\ref{fig:distance_vs_k}, which displays the $k$ values of the clustered stars versus their distance, reveals two different $k$ trends with distance for the two clusters. This provides further evidence that in these stars, the significant position angle rotation is caused by a change in the ISM between observer and source, and that the behavior of $k$ is strongly directional.

WR 25 has had a previous ISP estimate produced by \citet{drissen_polarization_1992}. They found $P_\mathrm{max}=6.74\pm0.02\%$ and $\lambda_\mathrm{max}=6050\pm10$\AA, using the standard Serkowksi law. Their $P_\mathrm{max}$ is identical to ours within uncertainties, though their $\lambda_\mathrm{max}$ is significantly smaller. This latter result is almost certainly due to the inclusion of $k$ in our fits. \citet{drissen_polarization_1992} noted that either there was a wavelength dependence of the ISP position angle or a wavelength-dependent intrinsic polarization of low magnitude. Since we find a significant $k$ value for WR 25, but no significant intrinsic polarization, it is likely that we have detected the proposed wavelength-dependent ISP position angle. \citet{drissen_polarization_1992} suggested that this could be due to the Carina nebula processing interstellar grains via shock waves. Our clustered $k$ values for WR 22, WR 23 and WR 25 support this conclusion, and we make the same suggestion as \citeauthor{drissen_polarization_1992}, that the Carina nebula could benefit from a concentrated ISP survey.


\subsection{Str\"omgren filter results}
\label{sec:stromgren}

The narrow Str\"omgren $b$ filter spans the complex $\lambda$4650 line region, which includes several strong emission lines in both WC and WN spectral types (Section~\ref{sec:fitting}). We used this filter to observe 19 stars in our sample. To determine the significance of our measurements, we calculated the residual of the $b$ filter data with respect to the fitted ISP equation in $q$, $u$ and $p$, including the intrinsic polarization if present, by subtracting the fit results from the $b$ filter data. We present the results in Table~\ref{tab:bfilter}. We considered the residual to be significant if its absolute value was 3$\sigma$ or more greater than the uncertainty on the measurement. Following the arguments by \citet{vink17wolf}, a $b$ filter measurement that is depolarized in $p$ compared to the intrinsic continuum polarization may be evidence of the line effect and thus imply an asymmetric WR wind and a rapidly rotating WR star. In the binary systems for which we could not define a systemic mean polarization, binary illumination of the WR wind may also contribute to the intrinsic continuum polarization.

\begin{deluxetable*}{lDDDD}
\label{tab:bfilter}
\tablecaption{Polarimetric residuals of our Str\"omgren $b$ filter observations with respect to the ISP + intrinsic fits presented in Table~\ref{tab:snapshot_isp} (in the sense $b- \textrm{fit}$). We also list the uncertainty on each $b$ measurement for comparison. 
}
\tablehead{WR  & \multicolumn2c{$b_q$ residual (\%)} & \multicolumn2c{$b_u$ residual (\%)} & \multicolumn2c{$b_p$ residual (\%)} & \multicolumn2c{$\sigma_b$ (\%)}}
\startdata
\decimals
6   & -0.045\tablenotemark{a} & 0.184\tablenotemark{a}  & -0.169\tablenotemark{a} & 0.009 \\
16  & -0.029 & 0.030  & 0.004  & 0.028 \\
21  & 0.061  & -0.030 & -0.022 & 0.058 \\
22  & -0.085 & 0.005  & 0.073  & 0.043 \\
24  & 0.052  & 0.054  & -0.074 & 0.037 \\
25  & 0.022  & -0.027 & 0.026  & 0.063 \\
40  & 0.001  & 0.024  & -0.020 & 0.025 \\
42  & 0.010  & 0.009  & -0.013 & 0.011 \\
48  & -0.017 & -0.108\tablenotemark{a} & -0.017 & 0.023 \\
52  & -0.015 & 0.018  & 0.017  & 0.042 \\
57  & 0.007  & 0.106  & 0.043  & 0.045 \\
69  & -0.072 & 0.010  & 0.043  & 0.068 \\
71  & -0.065 & 0.065  & 0.088  & 0.067 \\
78  & 0.021  & 0.037  & 0.041  & 0.024 \\
79  & 0.027\tablenotemark{a}  & -0.019 & -0.009 & 0.009 \\
90  & -0.441\tablenotemark{a} & 0.333\tablenotemark{a}  & 0.061  & 0.066 \\
103 & -0.080 & 0.016  & -0.034 & 0.034 \\
111 & 0.016  & 0.013  & -0.020 & 0.015 \\
113 & -0.157\tablenotemark{a} & 0.073  & 0.150\tablenotemark{a}  & 0.045
\enddata
\tablenotetext{a}{Denotes $|b|$ residual values $>3\sigma_b$.}
\tablecomments{The $b_p$ residual corresponds to a magnitude difference only, not a vector difference as with the $b_q$ and $b_u$ residuals. A negative $b_p$ value as defined here thus does not necessarily imply a depolarization in $q-u$ space.}
\end{deluxetable*}

Five stars showed a significant $b$ filter residual in any Stokes parameter: WR 6, WR 48, WR 79, WR 90, and WR 113. We checked each residual in $q-u$ space to verify whether it corresponded to a depolarization or a polarization enhancement with respect to the fit result. Three objects with significant $b$ residuals are known binaries: WR 48, WR 79 and WR 113. None of these binaries have a significant intrinsic continuum polarization using our 2$\sigma$ significance criterion (\S~\ref{sec:fitting}). In these cases, the residual in the $b$ filter may point to the existence of intrinsic continuum polarization (that was not sufficiently significant compared to our fitting uncertainties) at least equal to the $b$ residual value. We discuss each of these cases in more detail below.

WR 48 was only observed twice; it has a significant $b$ residual only in Stokes $u$, although we caution that because this result refers to a mean of two observations (\S~\ref{sec:binaryfit}f), the position angle of the residual is not well constrained. Given our uncertainties, this does not correspond to a significant line depolarization in $p$. Nonetheless, it does suggest some intrinsic continuum polarization, which may be due to an asymmetric WR wind, binary scattering effects, or both. 
Alternatively, WR 48 is a triple system whose third star, a O9.7Iab blue supergiant (BSG) is $\sim10\times$ brighter than the WR + O binary, so it is possible that the BSG is the source of the polarization, although this is rare amongst BSGs. This matches the findings of \citet{st.-louis_polarization_1987}, who detected stochastic, quasi-periodic fluctuations in the polarization of the system that they attributed to the O9.7Iab star.

WR 79 had 6 observations, so its $b$ uncertainty is low (\S~\ref{sec:binaryfit}e) and its $b_q$ residual is significant despite being small. As in WR 48, this is not a robust line effect detection, but it could indicate a slightly asymmetric WR wind. In addition, \citet{hill_modelling_2000} detected a wind collision region that could be asymmetric enough to produce intrinsic continuum polarization in this system that was not significant given our 2$\sigma$ criterion (\S~\ref{sec:fitting}). Additional phased observations at higher precision could further illuminate the nature of this continuum polarization.

WR 113 was observed twice (\S~\ref{sec:binaryfit}f) and thus, although its residual is significant in $q$ and not $u$, the same position angle caveat applies as in the case of WR 48. However, its $p$ residual is significant and positive. We verified in $q-u$ space that this residual is not a depolarization typical of the line effect, but rather an additional polarization in the $b$ filter in excess of our ISP + continuum fit. This implies that instead of being unpolarized, the $\lambda$4650 line region contains its own constant or varying intrinsic polarization, a result that may complicate studies of the line effect in some binary systems. Time-dependent spectropolarimetry is required to assess this possibility.


As noted in Section~\ref{sec:intrinsicpoldiscussion}, the binary status of WR 6 is ambiguous. The periodic nature of its polarization could be explained by the presence of CIRs or by a companion creating CIR-like structures in the wind \citep{harries_interstellar_1999}. Such structures could also give rise to the significant intrinsic polarization we detect (\S~\ref{sec:intrinsicpoldiscussion}). \citet{harries_interstellar_1999} also found that the region covered by the $b$ filter shows strong depolarization of the emission lines. Our negative $b_p$ residual, which again we verified in $q-u$ space supports this line effect detection. 
This depolarization has also affected the $B$ filter in our data, especially in Stokes $u$.  


WR 90 is particularly interesting because it displays an intrinsic polarization with greater than 5$\sigma$ significance, along with the significant Str\"omgren $b$ filter residual. The residual shows a rotation of the polarization position angle of 71.5$\degr$ with respect to the continuum, but no depolarization in $p$. Because this star has a WC7 spectral type, this deviation from the continuum angle is likely due to polarization effects in the \ion{C}{3} $\lambda$4650 line region. This may be the first evidence that WR 90 has a structured or aspherical wind with a preferred orientation angle. 
However, a study by \citet{chene_systematic_2011} showed only small-scale spectral variability, characteristic of clumps in the wind, without any sign of large-scale variability that could be attributed to the presence of a global wind structure. This may hint at transient structures, such as CIRs, in the WR 90 wind.

All five of the stars we found to contain a significant $b$ residual would benefit from focused, time-dependent spectropolarimetric observing campaigns to provide more information about the emission line polarization and reveal more details about the structures of their winds and other circumstellar material.

\section{Conclusions}

We observed a sample of 47 WR systems, both single and binary, using broadband $UBVRIb$ filter polarimetry. 
We fit a modified Serkowski law to the data to characterise each star's intrinsic polarization and ISP contribution. We provide a table of fitted ISP values (Table~\ref{tab:snapshot_isp}) and a sky map of ISP vectors (Figure~\ref{fig:isp_skymap}) as a resource for future polarimetric observations of these stars.

We found that 10 of the systems exhibit significant intrinsic polarization. Three of these stars (WR 21, WR 24 and WR 155) are short-period binaries and so their intrinsic polarization can be attributed to a combination of asymmetric winds due to rapid rotation of the WR star, illumination of the WR wind by the O star companion, and wind asymmetries caused by binary interaction. The intrinsic polarization in the other 7 systems is likely due to either complex wind structures (WR 6, WR 90, WR 134) or wind clumping (WR 14, WR 23, WR 103, WR 128), though WR 6 may have a binary companion. Six stars showed intrinsic polarization at 2--3$\sigma$ significance, and we suggest further observations of these targets to improve the uncertainties. Table~\ref{tab:limits} presents $1\sigma$ upper limits to the intrinsic polarization for all other stars to guide 
future observations.

Fourteen stars in our sample showed a significant wavelength dependence of the ISP position angle. Some of these objects are clustered closely on the sky, suggesting that the wavelength dependence is due to the effects of multiple dust clouds along the line of sight. We also confirm the result of \citet{bagnulo_large_2017} that large $|k|$ values have a weak inverse relationship with $P_\mathrm{IS,max}$ (Fig.~\ref{fig:pmax_vs_k}). 

Nineteen systems were observed with the Str\"omgren $b$ filter to investigate the $\lambda4650$ line complex present in most WR stars (Table~\ref{tab:bfilter}). Five stars showed a significant residual in the $b$ filter: WR 6, WR 48, WR 79, WR 90, and WR 113. Three of these are binaries (WR 48, WR 79 and WR 113). The residual of WR 48 may be due to a combination of effects, including an asymmetric wind collision region.
WR 79 is likely to have a wind collision region whose asymmetry contributes to the intrinsic polarization of the WR wind. WR 113 exhibits possible intrinsic line polarization, which is unusual and warrants further study. WR 6 has an ambiguous nature, so its residual could be explained either by CIR structures in its wind or by the motion of a binary companion creating structures in the wind. WR 90 is an interesting case, whose significant intrinsic polarization and  position angle rotation in the $b$ filter may indicate hitherto unknown asymmetries or structure in the wind.

We are currently monitoring 10 of the WR binary systems from this sample using spectropolarimetric observations obtained with the Robert Stobie Spectrograph on the Southern African Large Telescope \citep{fullard_spectropolarimetry_2018, johnson_comparison_2019}. These wavelength- and time-dependent data will enable us to characterize the colliding wind geometries and other binary properties in greater detail than has previously been possible. Similar observing campaigns focused on the other objects of interest highlighted here will reveal valuable information about the nature and structure of their WR winds. 

\acknowledgments
The authors thank the referee for insightful comments that improved this manuscript, and the assistance provided by the AAS statistical review team. Thanks to M. Shrestha for her helpful suggestions for improving Table~\ref{tab:starlist}.

AGF and JLH acknowledge support from NSF award AST-1816944.
AFJM and NSL are grateful to NSERC (Canada) for financial aid.

AGF and JLH respectfully acknowledge the Arapaho and Cheyenne peoples, upon whose traditional land the University of Denver was built. AGF and JLH also acknowledge the custodians and administrative staff at the University of Denver for their support. 

This research has made use of the SIMBAD database, operated at CDS, Strasbourg, France. The original description of the VizieR service was published in 2000, A\&AS 143, 9.

This research has made use of the VizieR catalogue access tool, CDS, Strasbourg, France (DOI: 10.26093/cds/vizier). The original description of the VizieR service was published in 2000, A\&AS 143, 23. 

%

\vspace{5mm}
\facilities{CrAO:1.25m, Danish 1.54m Telescope}


\software{astropy \citep{robitaille_astropy:_2013, the_astropy_collaboration_astropy_2018},
        emcee, \citep{foreman-mackey_emcee:_2013},
        lmfit \citep{newville_lmfit:_2014},
        matplotlib \citep{hunter_matplotlib:_2007},
        scipy \citep{virtanen_scipy_2019},
        seaborn (https://seaborn.pydata.org/)
          }



\appendix

\section{Polarimetric data for stars with multiple observations}

We present the data that were used in Section~\ref{sec:binaryfit} in Tables~\ref{tab:multiobsstars},~\ref{tab:wr133data} and~\ref{tab:wr134data}.

\begin{deluxetable*}{lDDDDDDDDDDDD}
\label{tab:multiobsstars}
\tablecaption{Observational data for objects with fewer than 5 observations. All objects in this table were observed at ESO/La Silla. }
\tablehead{\colhead{HJD ($UBVR$)} & \multicolumn2c{} & \multicolumn2c{$U$} & \multicolumn2c{} & \multicolumn2c{} & \multicolumn2c{$B$} & \multicolumn2c{} & \multicolumn2c{} & \multicolumn2c{$V$} & \multicolumn2c{} & \multicolumn2c{} & \multicolumn2c{$R$} & \multicolumn2c{} \\
\colhead{2,440,000+} & \multicolumn2c{$q$ (\%)} & \multicolumn2c{$u$ (\%)} & \multicolumn2c{$\sigma_p$ (\%)} & \multicolumn2c{$q$ (\%)} & \multicolumn2c{$u$ (\%)} & \multicolumn2c{$\sigma_p$ (\%)} & \multicolumn2c{$q$ (\%)} & \multicolumn2c{$u$ (\%)} & \multicolumn2c{$\sigma_p$ (\%)} & \multicolumn2c{$q$ (\%)} & \multicolumn2c{$u$ (\%)} & \multicolumn2c{$\sigma_p$ (\%)}}
\tabletypesize{\scriptsize}
\startdata
\decimals
WR 16     & .      & .      & .     & .      & .      & .     & .      & .      & .     & .      & .      & .    \\
\hline
412.5653 & -1.213 & -0.892 & 0.052 & -1.120 & -1.118 & 0.058 & -1.010 & -1.121 & 0.081 & -1.039 & -0.852 & 0.165 \\
413.5327 & -1.144 & -0.771 & 0.029 & -1.216 & -1.108 & 0.018 & -1.453 & -1.161 & 0.082 & -1.041 & -1.279 & 0.024 \\
415.5219 & -1.234 & -0.657 & 0.022 & -1.304 & -0.912 & 0.023 & -1.527 & -1.134 & 0.043 & -1.221 & -1.075 & 0.018 \\
417.5685 & -1.299 & -0.685 & 0.031 & -1.364 & -1.011 & 0.024 & -1.508 & -1.179 & 0.048 & -1.169 & -1.211 & 0.013 
\enddata
\tablecomments{Table~\ref{tab:multiobsstars} is published in its entirety in the machine-readable format. A portion is shown here for guidance regarding its form and content. Str\"omgren $b$ filter data are provided for some targets in the machine-readable table.}
\end{deluxetable*}

\begin{deluxetable*}{DDDDDDDDDDDDD}
\label{tab:wr133data}
\tablecaption{Polarimetric observations of WR 133. Observed at the Crimean Observatory.}
\tablehead{\multicolumn2c{HJD} & \multicolumn2c{} & \multicolumn2c{$U$} & \multicolumn2c{} & \multicolumn2c{} & \multicolumn2c{} & \multicolumn2c{$B$} & \multicolumn2c{} & \multicolumn2c{} & \multicolumn2c{} & \multicolumn2c{$V$} & \multicolumn2c{} & \multicolumn2c{} \\
\multicolumn2c{2,447,000+} & \multicolumn2c{$q$ (\%)} & \multicolumn2c{$\sigma_q$} & \multicolumn2c{$u$} & \multicolumn2c{$\sigma_u$ (\%)} & \multicolumn2c{$q$ (\%)} & \multicolumn2c{$\sigma_q$ (\%)} & \multicolumn2c{$u$ (\%)} & \multicolumn2c{$\sigma_u$ (\%)} & \multicolumn2c{$q$ (\%)} & \multicolumn2c{$\sigma_q$ (\%)} & \multicolumn2c{$u$ (\%)} & \multicolumn2c{$\sigma_u$ (\%)}}
\tabletypesize{\scriptsize}
\startdata
\decimals
320.408 & 0.446  & 0.025 & -0.310 & 0.030 & 0.386 & 0.048 & -0.349 & 0.013 & 0.295 & 0.021 & -0.321 & 0.035 \\
321.460 & 0.511  & 0.031 & -0.307 & 0.017 & 0.466 & 0.020 & -0.350 & 0.020 & 0.467 & 0.026 & -0.341 & 0.032 \\
322.453 & 0.160  & 0.047 & -0.248 & 0.089 & 0.233 & 0.061 & -0.322 & 0.061 & 0.039 & 0.096 & -0.486 & 0.074 \\
325.395 & 0.192  & 0.081 & -0.325 & 0.058 & 0.265 & 0.069 & -0.287 & 0.029 & 0.242 & 0.031 & -0.225 & 0.040 \\
329.412 & 0.352  & 0.105 & -0.353 & 0.057 & 0.329 & 0.052 & -0.350 & 0.006 & 0.237 & 0.040 & -0.350 & 0.024 
\enddata
\tablecomments{Table~\ref{tab:wr133data} is published in its entirety in the machine-readable format. A portion is shown here for guidance regarding its form and content. $R$ and $I$ filter data are provided in the machine-readable table.}
\end{deluxetable*}

\begin{deluxetable*}{DDDDDDDDDDDDD}
\label{tab:wr134data}
\tablecaption{Polarimetric observations of WR 134. Observed at the Crimean Observatory.}
\tablehead{\multicolumn2c{HJD} & \multicolumn2c{} & \multicolumn2c{$U$} & \multicolumn2c{} & \multicolumn2c{} & \multicolumn2c{} & \multicolumn2c{$B$} & \multicolumn2c{} & \multicolumn2c{} & \multicolumn2c{} & \multicolumn2c{$V$} & \multicolumn2c{} & \multicolumn2c{} \\
\multicolumn2c{2,447,700+} & \multicolumn2c{$q$ (\%)} & \multicolumn2c{$\sigma_q$} & \multicolumn2c{$u$} & \multicolumn2c{$\sigma_u$ (\%)} & \multicolumn2c{$q$ (\%)} & \multicolumn2c{$\sigma_q$ (\%)} & \multicolumn2c{$u$ (\%)} & \multicolumn2c{$\sigma_u$ (\%)} & \multicolumn2c{$q$ (\%)} & \multicolumn2c{$\sigma_q$ (\%)} & \multicolumn2c{$u$ (\%)} & \multicolumn2c{$\sigma_u$ (\%)}}
\tabletypesize{\scriptsize}
\startdata
\decimals
59.3018 & 1.204 & 0.048 & 0.358 & 0.079 & 0.946 & 0.032 & 0.218 & 0.048 & 1.091 & 0.062 & 0.303 & 0.068 \\
60.4033 & 1.051 & 0.034 & 0.265 & 0.025 & 0.872 & 0.039 & 0.179 & 0.028 & 0.912 & 0.040 & 0.203 & 0.026 \\
60.4692 & 0.946 & 0.070 & 0.395 & 0.060 & 0.790 & 0.050 & 0.229 & 0.054 & 0.858 & 0.046 & 0.268 & 0.068 \\
61.2734 & 1.182 & 0.059 & 0.441 & 0.104 & 0.990 & 0.034 & 0.256 & 0.077 & 1.015 & 0.039 & 0.275 & 0.081 \\
61.3306 & 1.360 & 0.108 & 0.340 & 0.089 & 1.145 & 0.044 & 0.244 & 0.065 & 1.261 & 0.051 & 0.176 & 0.071
\enddata
\tablecomments{Table~\ref{tab:wr134data} is published in its entirety in the machine-readable format. A portion is shown here for guidance regarding its form and content. $R$ and $I$ filter data are provided in the machine-readable table.}
\end{deluxetable*}

\section{Interstellar and intrinsic polarization fits}

\figsetstart
\figsetnum{8}
\figsettitle{Intrinsic and interstellar polarization fit results}

\figsetgrpstart
\figsetgrpnum{8.1}
\figsetgrptitle{WR 1 $UBVRI$ ISP fit.
}
\figsetplot{f1A_1.pdf}
\figsetgrpnote{$UBVRI$ data (black points) fit with equations \ref{eqn:qispshort} and \ref{eqn:uispshort} (green curves).}
\figsetgrpend

\figsetgrpstart
\figsetgrpnum{8.2}
\figsetgrptitle{WR 3 $UBVRI$ ISP fit.
}
\figsetplot{f1A_3.pdf}
\figsetgrpnote{$UBVRI$ data (black points) fit with equations \ref{eqn:qispshort} and \ref{eqn:uispshort} (green curves).}
\figsetgrpend

\figsetgrpstart
\figsetgrpnum{8.3}
\figsetgrptitle{WR 6 $UBVR$ ISP fit.
}
\figsetplot{f1A_6.pdf}
\figsetgrpnote{$UBVRI$ data (black points) fit with equations \ref{eqn:qisp} and \ref{eqn:uisp} (green curves). $b$ filter data are presented as blue points, but were not included as part of the fit.}
\figsetgrpend

\figsetgrpstart
\figsetgrpnum{8.4}
\figsetgrptitle{WR 8 $UBVR$ ISP fit.
}
\figsetplot{f1A_8.pdf}
\figsetgrpnote{$UBVRI$ data (black points) fit with equations \ref{eqn:qispshort} and \ref{eqn:uispshort} (green curves).}
\figsetgrpend

\figsetgrpstart
\figsetgrpnum{8.5}
\figsetgrptitle{WR 9 $UBVR$ ISP fit.
}
\figsetplot{f1A_9.pdf}
\figsetgrpnote{$UBVRI$ data (black points) fit with equations \ref{eqn:qispshort} and \ref{eqn:uispshort} (green curves).}
\figsetgrpend

\figsetgrpstart
\figsetgrpnum{8.6}
\figsetgrptitle{WR 14 $UBVR$ ISP fit.
}
\figsetplot{f1A_14.pdf}
\figsetgrpnote{$UBVRI$ data (black points) fit with equations \ref{eqn:qisp}, \ref{eqn:uisp} (green curves).}
\figsetgrpend

\figsetgrpstart
\figsetgrpnum{8.7}
\figsetgrptitle{WR 16 $UBVR$ ISP fit.
}
\figsetplot{f1A_16.pdf}
\figsetgrpnote{$UBVRI$ data (black points) fit with equations \ref{eqn:qisp} and \ref{eqn:uisp} (green curves). $b$ filter data are presented as blue points, but were not included as part of the fit.}
\figsetgrpend

\figsetgrpstart
\figsetgrpnum{8.8}
\figsetgrptitle{WR 21 $UBVR$ ISP fit.
}
\figsetplot{f1A_21.pdf}
\figsetgrpnote{$UBVRI$ data (black points) fit with equations \ref{eqn:qisp} and \ref{eqn:uisp} (green curves).  $b$ filter data are present as blue points, but were not included as part of the fit.}
\figsetgrpend

\figsetgrpstart
\figsetgrpnum{8.9}
\figsetgrptitle{WR 22 $UBVR$ ISP fit.
}
\figsetplot{f1A_22.pdf}
\figsetgrpnote{$UBVRI$ data (black points) fit with equations \ref{eqn:qisp} and \ref{eqn:uisp} (green curves). $b$ filter data are present as blue points, but were not included as part of the fit.}
\figsetgrpend

\figsetgrpstart
\figsetgrpnum{8.10}
\figsetgrptitle{WR 23 $UBVR$ ISP fit.
}
\figsetplot{f1A_23.pdf}
\figsetgrpnote{$UBVRI$ data (black points) fit with equations \ref{eqn:qisp} and \ref{eqn:uisp} (green curves).}
\figsetgrpend

\figsetgrpstart
\figsetgrpnum{8.11}
\figsetgrptitle{WR 24 $UBVR$ ISP fit.
}
\figsetplot{f1A_24.pdf}
\figsetgrpnote{$UBVRI$ data (black points) fit with equations \ref{eqn:qisp} and \ref{eqn:uisp} (green curves). $b$ filter data are present as blue points, but were not included as part of the fit.}
\figsetgrpend

\figsetgrpstart
\figsetgrpnum{8.12}
\figsetgrptitle{WR 25 $UBVR$ ISP fit.
}
\figsetplot{f1A_25.pdf}
\figsetgrpnote{$UBVRI$ data (black points) fit with equations \ref{eqn:qispshort} and \ref{eqn:uispshort} (green curves). $b$ filter data are present as blue points, but were not included as part of the fit.}
\figsetgrpend

\figsetgrpstart
\figsetgrpnum{8.13}
\figsetgrptitle{WR 40 $UBVR$ ISP fit.
}
\figsetplot{f1A_40.pdf}
\figsetgrpnote{$UBVRI$ data (black points) fit with equations \ref{eqn:qispshort} and \ref{eqn:uispshort} (green curves). $b$ filter data are present as blue points, but were not included as part of the fit.}
\figsetgrpend

\figsetgrpstart
\figsetgrpnum{8.14}
\figsetgrptitle{WR 42 $UBVR$ ISP fit.
}
\figsetplot{f1A_42.pdf}
\figsetgrpnote{$UBVRI$ data (black points) fit with equations \ref{eqn:qisp} and \ref{eqn:uisp} (green curves). $b$ filter data are present as blue points.}
\figsetgrpend

\figsetgrpstart
\figsetgrpnum{8.15}
\figsetgrptitle{WR 43 $UBVR$ ISP fit.
}
\figsetplot{f1A_43.pdf}
\figsetgrpnote{$UBVRI$ data (black points) fit with equations \ref{eqn:qispshort} and \ref{eqn:uispshort} (green curves).}
\figsetgrpend

\figsetgrpstart
\figsetgrpnum{8.16}
\figsetgrptitle{WR 46 $UBVR$ ISP fit.
}
\figsetplot{f1A_46.pdf}
\figsetgrpnote{$UBVRI$ data (black points) fit with equations \ref{eqn:qispshort} and \ref{eqn:uispshort} (green curves).}
\figsetgrpend

\figsetgrpstart
\figsetgrpnum{8.17}
\figsetgrptitle{WR 48 $UBVR$ ISP fit.
}
\figsetplot{f1A_48.pdf}
\figsetgrpnote{$UBVRI$ data (black points) fit with equations \ref{eqn:qispshort} and \ref{eqn:uispshort} (green curves). $b$ filter data are present as blue points, but were not included as part of the fit.}
\figsetgrpend

\figsetgrpstart
\figsetgrpnum{8.18}
\figsetgrptitle{WR 52 $UBVR$ ISP fit.
}
\figsetplot{f1A_52.pdf}
\figsetgrpnote{$UBVRI$ data (black points) fit with equations \ref{eqn:qispshort} and \ref{eqn:uispshort} (green curves). $b$ filter data are present as blue points, but were not included as part of the fit.}
\figsetgrpend

\figsetgrpstart
\figsetgrpnum{8.19}
\figsetgrptitle{WR 57 $UBVR$ ISP fit.
}
\figsetplot{f1A_57.pdf}
\figsetgrpnote{$UBVRI$ data (black points) fit with equations \ref{eqn:qispshort} and \ref{eqn:uispshort} (green curves). $b$ filter data are present as blue points, but were not included as part of the fit.}
\figsetgrpend

\figsetgrpstart
\figsetgrpnum{8.20}
\figsetgrptitle{WR 69 $UBVR$ ISP fit.
}
\figsetplot{f1A_69.pdf}
\figsetgrpnote{$UBVRI$ data (black points) fit with equations \ref{eqn:qispshort} and \ref{eqn:uispshort} (green curves). $b$ filter data are present as blue points, but were not included as part of the fit.}
\figsetgrpend

\figsetgrpstart
\figsetgrpnum{8.21}
\figsetgrptitle{WR 71 $UBVR$ ISP fit.
}
\figsetplot{f1A_71.pdf}
\figsetgrpnote{$UBVRI$ data (black points) fit with equations \ref{eqn:qisp} and \ref{eqn:uisp} (green curves). $b$ filter data are present as blue points, but were not included as part of the fit.}
\figsetgrpend

\figsetgrpstart
\figsetgrpnum{8.22}
\figsetgrptitle{WR 78 $UBVR$ ISP fit.
}
\figsetplot{f1A_78.pdf}
\figsetgrpnote{$UBVRI$ data (black points) fit with equations \ref{eqn:qispshort} and \ref{eqn:uispshort} (green curves). $b$ filter data are present as blue points, but were not included as part of the fit.}
\figsetgrpend

\figsetgrpstart
\figsetgrpnum{8.23}
\figsetgrptitle{WR 79 $UBVR$ ISP fit.
}
\figsetplot{f1A_79.pdf}
\figsetgrpnote{$UBVRI$ data (black points) fit with equations \ref{eqn:qispshort} and \ref{eqn:uispshort} (green curves). $b$ filter data are present as blue points, but were not included as part of the fit.}
\figsetgrpend

\figsetgrpstart
\figsetgrpnum{8.24}
\figsetgrptitle{WR 86 $UBVR$ ISP fit.
}
\figsetplot{f1A_86.pdf}
\figsetgrpnote{$UBVRI$ data (black points) fit with equations \ref{eqn:qispshort} and \ref{eqn:uispshort} (green curves).}
\figsetgrpend

\figsetgrpstart
\figsetgrpnum{8.25}
\figsetgrptitle{WR 90 $UBVR$ ISP fit.
}
\figsetplot{f1A_90.pdf}
\figsetgrpnote{$UBVRI$ data (black points) fit with equations \ref{eqn:qisp} and \ref{eqn:uisp} (green curves). $b$ filter data are present as blue points, but were not included as part of the fit.}
\figsetgrpend

\figsetgrpstart
\figsetgrpnum{8.26}
\figsetgrptitle{WR 92 $UBVR$ ISP fit.
}
\figsetplot{f1A_92.pdf}
\figsetgrpnote{$UBVRI$ data (black points) fit with equations \ref{eqn:qispshort} and \ref{eqn:uispshort} (green curves).}
\figsetgrpend

\figsetgrpstart
\figsetgrpnum{8.27}
\figsetgrptitle{WR 103 $UBVR$ ISP fit.
}
\figsetplot{f1A_103.pdf}
\figsetgrpnote{$UBVRI$ data (black points) fit with equations \ref{eqn:qisp} and \ref{eqn:uisp} (green curves). $b$ filter data are present as blue points, but were not included as part of the fit.}
\figsetgrpend

\figsetgrpstart
\figsetgrpnum{8.28}
\figsetgrptitle{WR 108 $UBVR$ ISP fit.
}
\figsetplot{f1A_108.pdf}
\figsetgrpnote{$UBVRI$ data (black points) fit with equations \ref{eqn:qisp} and \ref{eqn:uisp} (green curves).}
\figsetgrpend

\figsetgrpstart
\figsetgrpnum{8.29}
\figsetgrptitle{WR 110 $UBVR$ ISP fit.
}
\figsetplot{f1A_110.pdf}
\figsetgrpnote{$UBVRI$ data (black points) fit with equations \ref{eqn:qispshort} and \ref{eqn:uispshort} (green curves).}
\figsetgrpend

\figsetgrpstart
\figsetgrpnum{8.30}
\figsetgrptitle{WR 111 $UBVR$ ISP fit.
}
\figsetplot{f1A_111.pdf}
\figsetgrpnote{$UBVRI$ data (black points) fit with equations \ref{eqn:qisp} and \ref{eqn:uisp} (green curves). $b$ filter data are present as blue points, but were not included as part of the fit.}
\figsetgrpend

\figsetgrpstart
\figsetgrpnum{8.31}
\figsetgrptitle{WR 113 $UBVR$ ISP fit.
}
\figsetplot{f1A_113.pdf}
\figsetgrpnote{$UBVRI$ data (black points) fit with equations \ref{eqn:qispshort} and \ref{eqn:uispshort} (green curves). $b$ filter data are present as blue points, but were not included as part of the fit.}
\figsetgrpend

\figsetgrpstart
\figsetgrpnum{8.32}
\figsetgrptitle{WR 123 $UBVR$ ISP fit.
}
\figsetplot{f1A_123.pdf}
\figsetgrpnote{$UBVRI$ data (black points) fit with equations \ref{eqn:qispshort} and \ref{eqn:uispshort} (green curves).}
\figsetgrpend

\figsetgrpstart
\figsetgrpnum{8.33}
\figsetgrptitle{WR 127 $UBVRI$ ISP fit.
}
\figsetplot{f1A_127.pdf}
\figsetgrpnote{$UBVRI$ data (black points) fit with equations \ref{eqn:qispshort} and \ref{eqn:uispshort} (green curves).}
\figsetgrpend

\figsetgrpstart
\figsetgrpnum{8.34}
\figsetgrptitle{WR 128 $UBVRI$ ISP fit.
}
\figsetplot{f1A_128.pdf}
\figsetgrpnote{$UBVRI$ data (black points) fit with equations \ref{eqn:qisp} and \ref{eqn:uisp} (green curves).}
\figsetgrpend

\figsetgrpstart
\figsetgrpnum{8.35}
\figsetgrptitle{WR 133 $UBVRI$ ISP fit.
}
\figsetplot{f1A_133.pdf}
\figsetgrpnote{$UBVRI$ data (black points) fit with equations \ref{eqn:qisp} and \ref{eqn:uisp} (green curves).}
\figsetgrpend

\figsetgrpstart
\figsetgrpnum{8.36}
\figsetgrptitle{WR 134 $UBVRI$ ISP fit.
}
\figsetplot{f1A_134.pdf}
\figsetgrpnote{$UBVRI$ data (black points) fit with equations \ref{eqn:qisp} and \ref{eqn:uisp} (green curves).}
\figsetgrpend

\figsetgrpstart
\figsetgrpnum{8.37}
\figsetgrptitle{WR 135 $UBVRI$ ISP fit.
}
\figsetplot{f1A_135.pdf}
\figsetgrpnote{$UBVRI$ data (black points) fit with equations \ref{eqn:qispshort} and \ref{eqn:uispshort} (green curves).}
\figsetgrpend

\figsetgrpstart
\figsetgrpnum{8.38}
\figsetgrptitle{WR 136 $UBVRI$ ISP fit.
}
\figsetplot{f1A_136.pdf}
\figsetgrpnote{$UBVRI$ data (black points) fit with equations \ref{eqn:qispshort} and \ref{eqn:uispshort} (green curves).}
\figsetgrpend

\figsetgrpstart
\figsetgrpnum{8.39}
\figsetgrptitle{WR 137 $UBVRI$ ISP fit.
}
\figsetplot{f1A_137.pdf}
\figsetgrpnote{$UBVRI$ data (black points) fit with equations \ref{eqn:qispshort} and \ref{eqn:uispshort} (green curves).}
\figsetgrpend

\figsetgrpstart
\figsetgrpnum{8.40}
\figsetgrptitle{WR 138 $UBVRI$ ISP fit.
}
\figsetplot{f1A_138.pdf}
\figsetgrpnote{$UBVRI$ data (black points) fit with equations \ref{eqn:qispshort} and \ref{eqn:uispshort} (green curves).}
\figsetgrpend

\figsetgrpstart
\figsetgrpnum{8.41}
\figsetgrptitle{WR 139 $UBVRI$ ISP fit.
}
\figsetplot{f1A_139.pdf}
\figsetgrpnote{$UBVRI$ data (black points) fit with equations \ref{eqn:qisp} and \ref{eqn:uisp} (green curves).}
\figsetgrpend

\figsetgrpstart
\figsetgrpnum{8.42}
\figsetgrptitle{WR 140 $UBVRI$ ISP fit.
}
\figsetplot{f1A_140.pdf}
\figsetgrpnote{$UBVRI$ data (black points) fit with equations \ref{eqn:qispshort} and \ref{eqn:uispshort} (green curves).}
\figsetgrpend

\figsetgrpstart
\figsetgrpnum{8.43}
\figsetgrptitle{WR 141 $UBVRI$ ISP fit.
}
\figsetplot{f1A_141.pdf}
\figsetgrpnote{$UBVRI$ data (black points) fit with equations \ref{eqn:qispshort} and \ref{eqn:uispshort} (green curves).}
\figsetgrpend

\figsetgrpstart
\figsetgrpnum{8.44}
\figsetgrptitle{WR 148 $UBVRI$ ISP fit.
}
\figsetplot{f1A_148.pdf}
\figsetgrpnote{$UBVRI$ data (black points) fit with equations \ref{eqn:qispshort} and \ref{eqn:uispshort} (green curves).}
\figsetgrpend

\figsetgrpstart
\figsetgrpnum{8.45}
\figsetgrptitle{WR 153 $UBVRI$ ISP fit.
}
\figsetplot{f1A_153.pdf}
\figsetgrpnote{$UBVRI$ data (black points) fit with equations \ref{eqn:qispshort} and \ref{eqn:uispshort} (green curves).}
\figsetgrpend

\figsetgrpstart
\figsetgrpnum{8.46}
\figsetgrptitle{WR 155 $UBVRI$ ISP fit.
}
\figsetplot{f1A_155.pdf}
\figsetgrpnote{$UBVRI$ data (black points) fit with equations \ref{eqn:qisp} and \ref{eqn:uisp} (green curves).}
\figsetgrpend

\figsetgrpstart
\figsetgrpnum{8.47}
\figsetgrptitle{WR 157 $UBVRI$ ISP fit.
}
\figsetplot{f1A_157.pdf}
\figsetgrpnote{$UBVRI$ data (black points) fit with equations \ref{eqn:qispshort} and \ref{eqn:uispshort} (green curves).}
\figsetgrpend

\figsetend

\begin{figure*}
\label{fig:figureset}
    \centering
    \includegraphics[width=\textwidth]{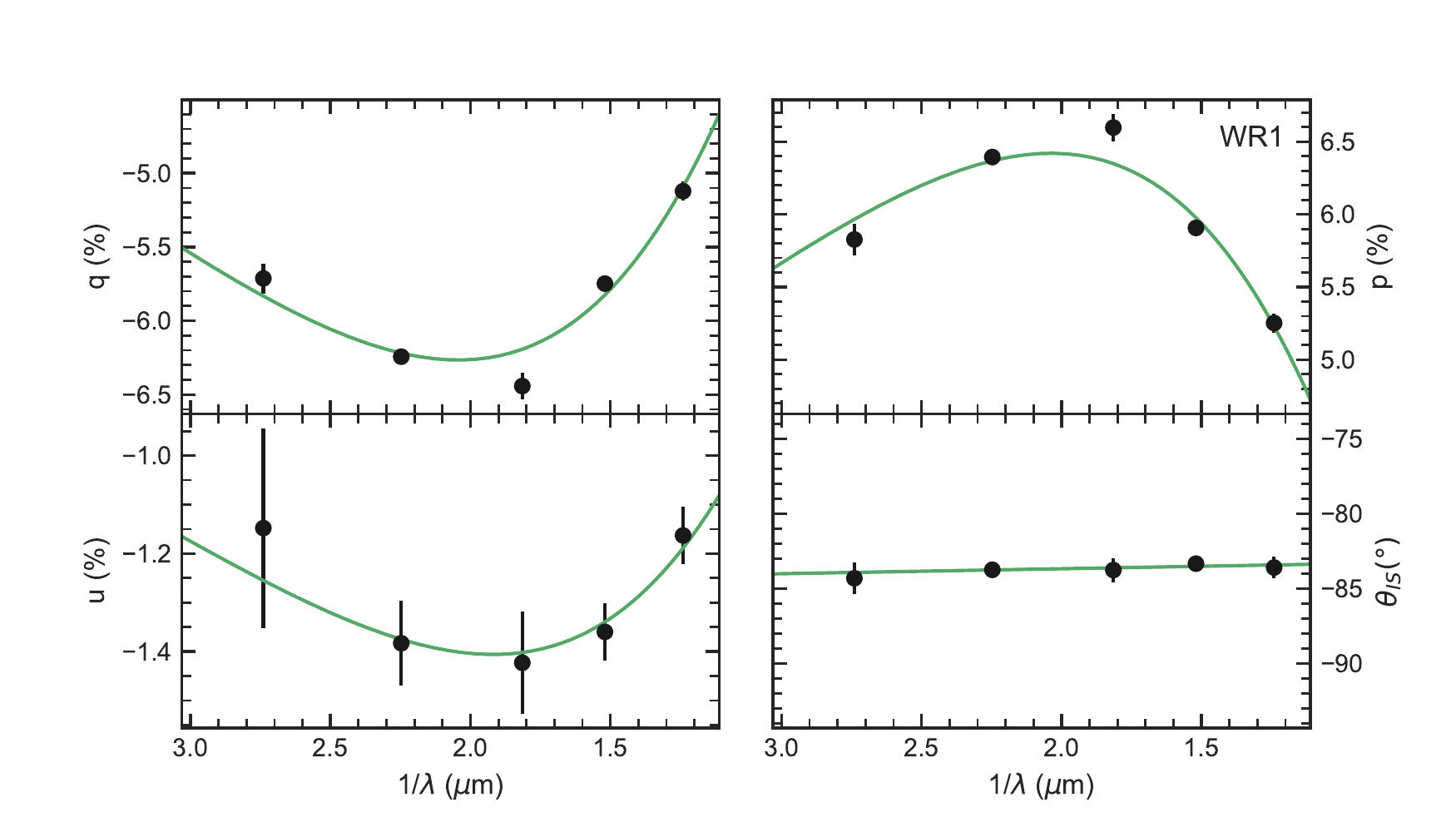}
\caption{The complete figure set (47 images) is available in the online journal.\\
WR 1 $UBVRI$ data (black points) fit with equations \ref{eqn:qispshort} and \ref{eqn:uispshort} (green curves).}
\end{figure*}

\bibliography{WR_ISP_paper}{}
\bibliographystyle{aasjournal}



\end{document}